\documentclass[traditabstract]{aa}
\usepackage[varg]{txfonts}
\usepackage{graphicx}
\usepackage{txfonts}
\usepackage{lipsum}
\usepackage{subcaption}         
\usepackage{lscape}             
\usepackage{placeins}           
\usepackage{natbib}
\usepackage{fancyhdr} 
\usepackage{amssymb}
\usepackage{xcolor}
\usepackage{caption}
\usepackage{afterpage}
\usepackage{dpfloat}
\usepackage{longtable}
\usepackage[colorlinks=true,
            linkcolor=blue,
            citecolor=blue,
            urlcolor=blue]{hyperref}

\newcommand\amm{NH$_{3}$}

\newcommand\nonV{$\sigma_{non}$}

\newcommand\kms{km s$^{-1}$}

\begin{document}

    \title{The gas kinematics  in 70 ${\mu}m$ dark molecular clumps with ammonia}


%

    \author{Chao Ou\inst{1}\fnmsep
        \and Shanghuo Li\inst{2,3}\fnmsep\thanks{shli@nju.edu.cn}
        \and Junzhi Wang\inst{1}\fnmsep\thanks{junzhiwang@gxu.edu.cn}
        \and Yuqiang Li\inst{4}
        \and Shuting Lin\inst{5}
        \and Ian W. Stephens\inst{6,7}
        \and Qizhou Zhang\inst{7}
        \and Bo Zhang\inst{8,9}
        }

   \institute{Guangxi Key Laboratory for Relativistic Astrophysics, School of Physical Science and Technology, Guangxi University, Nanning 530004, PR China
   \and School of Astronomy and Space Science, Nanjing University, 163 Xianlin Avenue, Nanjing 210023, People’s Republic of China
   \and Key Laboratory of Modern Astronomy and Astrophysics (Nanjing University), Ministry of Education, Nanjing 210023, People’s Republic of China
   \and Korea Astronomy and Space Science Institute, No. 776, Daedeok-daero, Yuseong-gu, Daejeon, Republic of Korea
   \and Department of Astronomy, Xiamen University, Zengcuo’an West Road, Xiamen 361005, China
   \and Department of Earth, Environment, and Physics, Worcester State University, Worcester, MA 01602, USA
   \and Center for Astrophysics | Harvard \& Smithsonian, 60 Garden Street, Cambridge, MA 02138, USA
   \and Shanghai Astronomical Observatory, Chinese Academy of Sciences, 80 Nandan Road, Shanghai 200030, China
   \and State Key Laboratory of Radio Astronomy and Technology, A20 Datun Road, Chaoyang District, Beijing, 100101, P. R. China
   }

   \date{Received XX, 2026}

   \abstract
    {We present the investigation of the gas kinematics and assess the evolutionary stages of dense cores embedded in four infrared dark clouds (IRDCs) using the  NH$_3$(1,1) and NH$_3$(2,2) lines obtained from the VLA and GBT observations. There is no 1.3 cm continuum emission counterpart in 1.3 mm continuum emission revealed by the SMA toward these IRDCs.
    The low production rate $N_{\rm uv} \sim 10^{44}$~s$^{-1}$ of Lyman continuum photons, indicates that the four IRDCs are in very early evolutionary stages, in which free--free emission is still absent.
    We have identified 61 dense cores at the size scale of $\sim$0.1 pc using the inner satellite groups of \amm(1,1) line. Among them, 38 dense cores exhibit a single velocity component, while 23 dense cores show multiple velocity components. We find that the nonthermal velocity dispersion ($\sigma_{\mathrm{non}}$) increases with increasing radial distance from the center of dense core within the inner 0.1\,pc toward eight dense cores, indicating that the turbulence is likely dissipated toward the center of these dense cores.
    In addition, two dense cores in AGAL~031.024+00.262 and one core in AGAL~024.314+00.086 exhibit nearly sonic motions on a small scale of 0.01 pc. The weakening of nonthermal support against gravity suggests that these dense cores  are close to gravitational collapse or are already collapsing.
    Conversely, there areseven dense cores in AGAL031.024+00.262 showing a decrease in $\sigma_{\mathrm{non}}$ with increasing radial distance from the center. By comparing this trends with the outflow revealed by CO J=2-1 line, we find that these cores are likely associated with embedded star formation activities. Higher angular resolution observations on sub-parsec scales are essential to reveal the transition from thermal to nonthermal-dominated regions, especially in dense cores associated with multiple velocity components.}


   \maketitle

\nolinenumbers
\section{Introduction}
\label{sec:intro}
\begin{table*}
\caption{\label{tab:table1}Summary of the observations.}
\centering
\setlength{\tabcolsep}{3pt}
\begin{tabular}{ccccccccccc}
\hline\hline
Date                  & Sources        & R.A.       & Decl.        & Cal. & Cal. & Cal. & N$_{ant}$ & D & $\theta_{\rm NH_3}$& $\theta_{\rm 1.3\,cm}$ \\
                      &                & (hh:mm:ss) & (dd:mm:ss)  & Flux       & Bandpass   & Phase      &           & (kpc)    & ($\arcsec$) &($\arcsec$ $\times$ $\arcsec$, $^{\circ}$)  \\
\hline
2018 Nov 01 & AGAL014.492–00.139 & 18:17:22.2 & -16:24:58.4 & 3C286 & J1743-0350 & J1832-1035 & 25 & 3.86 & 4.52 & 6.0$\times$3, 31.8\\
2018 Oct 19 & AGAL031.024+00.262 & 18:47:01.0 & -01:34:41.0 & 3C286 & J1743-0350 & J1832-1035 & 24 & 4.89 & 4.27 &4.1$\times$2.7, -0.5\\
2021 Apr 06  & AGAL031.024+00.262 & 18:47:01.0 & -01:34:41.0 & 3C286 & J1743-0350 & J1832-1035 & 27 &       &     &  \\
2021 Apr 08  & AGAL022.376+00.447 & 18:30:37.0 & -09:12:47.2 & 3C286 & J1743-0350 & J1832-1035 & 27 & 4.74  & 5.29& 5.6$\times$3.3, -46.0\\
2021 Mar 23 & AGAL024.314+00.086 & 18:35:19.0 & -07:37:27.2 & 3C286 & J1743-0350 & J1832-1035 & 25 & 6.54  & 5.89& 4.8$\times$2.8, 18.9\\
\hline
\end{tabular}
\tablefoot{Columns: (1) Observing date; (2) Source name; (3–4) Right ascension and declination; (5–7) Flux, bandpass, and phase calibrators; (8) Number of antennas; (9) Distance from the sun; (10) Circular beam size of \amm(1,1) data; (11) Beam size of 1.3\,cm continuum.}
\end{table*}

Star formation occurs through the collapse of dense cores within molecular clouds \citep{2007ARA&A..45..481Z,2018ARA&A..56...41M}. In the early stages, nonthermal motions or turbulence within dense cores play an important role by providing support against gravity \citep{2007ARA&A..45..565M}. However, once turbulence dissipates on small spatial scales of less than 0.1~pc, thermal motions can become the dominant source of support \citep{1998ApJ...507L.157M,2000ApJ...537..891W}. This transition accelerates the onset of gravitational collapse, making it crucial to determine the evolutionary phase of the dense cores.
To investigate turbulence dissipation, it is necessary to characterize the physical environment of dense cores, particularly their kinematic properties prior to star formation (e.g., \citealp{2014ApJ...789...83T,2016MNRAS.459.4130O,2020ApJ...896..110L}). Prestellar cores, the immediate precursors of protostellar cores, are deeply embedded within protoclusters, which makes them difficult to observe directly during very early stage \citep{2006ApJ...641..389R,2007ARA&A..45..339B}. 
Before the onset of collapse, the formation mechanisms of high-mass prestellar cores remain far less understood than those of low-mass cores, largely because of the limited observational constraints available for these earliest evolutionary stages \citep{2025ARA&A..63....1B}.

Infrared dark clouds (IRDCs) are considered potential sites for the early stage of star formation and have been extensively studied in recent decades \citep{2007ARA&A..45..339B, 2011ApJ...733...26Z, 2014ApJ...787..113B, 2019ApJ...886..130L, 2019ApJ...886..102S,2020ApJ...896..110L}. \cite{1998ApJ...494L.199E} first used the 8 $\mu$m dust data from the mid-IR survey conducted by the Midcourse Space Experiment (MSX) to identify numerous small, cold, and dense clouds. Building on this result, \cite{1998ApJ...508..721C} conducted rotational line observations of H$_{2}$CO to investigate the physical properties of ten MSX dark clouds, confirming temperature of 20 K and volume density of 10$^{5}$ cm$^{-3}$. \cite{2009A&A...505..405P} utilized the Spitzer GLIMPSE and MIPSGAL data to further identify Spitzer dark clouds, 80\% of which were previously undetected in the MSX dark clouds. \cite{2010A&A...518L.100M} used the Herschel data across five longer wavelengths obtained from the Hi-GAL survey. Based on the catalog by \cite{2009A&A...504..415S} using the Atacama Pathfinder Experiment (APEX) surveys towards the Galaxy at 870 $\mu$m, observations with high spatial resolution provided more detailed features of the clumps through interferometric arrays (e.g., \citealp{2019ApJ...886..102S}; \citealp{2021ApJ...923..263T}; \citealp{2024ApJ...966...54P}, etc.). Most research focuses on the fragmentation from parsec to subparsec scales towards massive clumps, offering a series of snapshots of different evolutionary stages of IRDCs \citep{2014MNRAS.439.3275W, 2019ApJ...886...36S, 2023MNRAS.526.2278A}. Furthermore, attention is given to the condensed regions of clumps that are perceived as prestellar cores. Systematic analysis of dense cores in IRDCs, including their kinematics and dynamics, has unveiled the initial conditions of the early stages through dust and molecular line emission \citep{2011ApJ...733...44D, 2015AJ....150..159D,2020ApJ...903..119L, 2023MNRAS.522.3357P,2023ApJ...949..109L}.

A recent study by \cite{2019ApJ...886..130L} reported the results of seven 70 $\mu$m IRDCs from SMA observations. Forty-four dense cores were identified by the 1.3 mm continuum and $^{12}$CO J=2-1 and C$^{18}$O J=2-1 were employed to study the dynamical property.
Nine dense cores have masses exceeding 27 M$_\odot$, potentially leading to the formation of massive stars in the protoclusters assuming 30\% of core star formation efficiency. CO outflows were detected in seventeen dense cores, which are distributed in 6 out of 7 clumps, indicating that these 70 $\mu$m dark clumps are not quiescent. The spectral lines of HCN J=3-2 and HCO$^{+}$ J=3-2 are simultaneously detected in compact regions (e.g., MM 1 and MM 7 in AGAL031.024+00.262).  
The C-bearing molecules generally freeze out on the surface of dust at temperatures below 20 K, leading to underestimated abundances \citep{2001ApJ...557..736G,2012A&ARv..20...56C, 2022ApJ...929...13C}.
However, N-bearing species such as NH$_3$ and N$_2$H$^+$ are largely unaffected in similar environments and are widely used to study the properties of dense cores \citep{2010A&A...513A..41H,2010ApJ...708.1002F,2017A&A...606A.123H}. 
The inversion line ratios of NH$_3$(1,1) and NH$_3$(2,2) have been commonly used in other clouds to determine the kinetic temperature of dense gas in cold regions ($\sim$15~K) \citep{1983ARA&A..21..239H,2004A&A...416..191T,2014ApJ...790...84L,2022A&A...667A.137B}.
Moreover, the intrinsic offset width of the hyperfine structure is not affected by feedback and turbulence, provided that it serves as a tracer for the kinematics of the clumps (e.g., \citealp[]{2012A&A...544A.146W, 2015A&A...579A..91W, 2020A&A...644A.128S}). In particular, the kinematic properties of ammonia in dense regions of IRDCs at early evolutionary stages provide a powerful diagnostic for identifying turbulence dissipation and for further confirming the gravitational collapse of dense cores leading to star formation \citep{1998ApJ...507L.157M, 2008ApJ...672L..33W}. To characterize the initial internal kinematics of dense cores and evaluate the role of nonthermal and thermal support in the early phases of dense cores, we use ammonia transitions to probe the kinematics of dense gas in  70~$\mu$m IRDCs.

We selected four sources from the original sample of seven presented in \citet{2019ApJ...886..130L} and updated their distances using the method established by \citet{2019ApJ...885..131R}. The remaining three 70~$\mu$m IRDCs were excluded because the VLA observations of AGAL016.418–00.634 suffer from significant missing flux and lack complementary GBT data, the VLA data for AGAL008.691–00.401 are of relatively poor quality, and AGAL030.844+00.177 was not observed in this work. 
In this study, we use the NH$_3$(1,1) and NH$_3$(2,2) data obtained from the combined VLA and GBT observations to investigate the kinematics of dense cores identified from the ammonia emission.
This paper is organized as follows. Sect.~\ref{sec:obs} describes the VLA and GBT observations and data reduction. Sect.~\ref{sec:met} presents the methods used for the analysis of the ammonia lines. Sect.~\ref{sec:result} presents the results from the continuum and ammonia line observations. The kinematic properties are discussed in Sect.~\ref{sec:disc}, and are summarized in Sect.~\ref{sec:col}.

\section{Observation and data reduction} \label{sec:obs}
\subsection{ Very Large Array observation}

VLA observation of the K band in the D-array configuration was carried out in 2018 November and October and in 2021 April, March, and June (Project code: 18A-250 and 21A-128, PI: Shanghuo Li). Ammonia inversion transitions NH$_{3}$(J,K)=(1,1) at 23.6944955 GHz and NH$_{3}$(J,K)=(2,2) at 23.7226333 GHz were simultaneously observed in four sources. The 8~MHz bandwidth consists of 1024 channels with a channel width of 7.812~kHz, corresponding to a velocity resolution of 0.1~km~s$^{-1}$ at 23.7~GHz in dual-polarization mode. For the continuum bandwidth of 128~MHz, there are 64 channels per 2~MHz. A total of 14 spectral windows (SPWs) were combined for the continuum, resulting in about 1.6 GHz  line-free bandwidth for imaging the 1.3 cm continuum.

The calibration and imaging of data are based on the CASA 6.5.4 software package \citep{2022PASP..134k4501C}. We use the same calibrator for every IRDC that was listed in Table \ref{tab:table1}. AGAL031.024+00.262 was observed in two projects that were combined to a uvdata with noise weighting. To achieve a high signal-to-noise ratio (S/N), we use natural weighting to image the 1.3 cm continuum. The beam parameters are provided in the last column of Table~\ref{tab:table1}.

\subsection{ GBT archive data}

The Radio Ammonia Mid-Plane Survey (RAMPS) was conducted between March~16,~2014 and January~31,~2019 using the 100~m Robert C.~Byrd Green Bank Telescope (GBT; \citealt{2018ApJS..237...27H}).
The pilot survey was mapping the first quadrant of the Galaxy (l = 10°–40°, $\vert{b}\vert$$<$0.4°), including four sources of VLA observation. The K-band Focal Plane Array (KFPA) consists of seven receivers covering a frequency range of 18–27.5~GHz. 
The observations were conducted with a spectral resolution corresponding to a channel width of 1.4~kHz, providing a velocity resolution of 0.018~km~s$^{-1}$ at the rest frequency of NH$_3$(1,1). 
At this frequency, each receiver has a beam size of approximately 34$\arcsec$, corresponding to the full width at half maximum (FWHM) of a Gaussian beam. 
Further details can be found in \cite{2018ApJS..237...27H}.

\subsection{Combine VLA data and GBT data}
\label{sec:combine}

To mitigate the missing-flux problem inherent to interferometric observations, we follow an established procedure that combines single-dish data and interferometer data \footnote{https://github.com/baobabyoo/almica}. The procedure is carried out in the uv domain that has pseudo visibility generated from the GBT cube using the MIRIAD software package and then joined to interferometric visibilities to the image \citep{1995ASPC...77..433S}. We were firstly using the \texttt{regrid} task to match the spatial resolution and spectral resolution of GBT data and VLA data. Making pseudo visibility was operated in \texttt{uvrandom} and \texttt{uvmodel} tasks after deconvolution for the GBT cube. And then, the inverse Fourier transform of the combined visibility in the \texttt{invert} task product a dirty image. 
We applied a robust weighting of 0.5 for AGAL014.492–00.139 and AGAL022.376+00.447 to effectively suppress sidelobe contamination, and a robust weighting of 1.5 for AGAL024.314+00.086 and AGAL031.024+00.262 to better recover extended emission during the \texttt{clean} task. The final images were produced using the \texttt{restor} task, which uses the clean beam to convolve the clean model.

After combining the VLA and GBT data, more than 90\% of the total flux in ammonia lines is recovered.
 For each IRDC, we smoothed the original elliptical synthesized beam along its major axis in order to obtain a circle, independent of beam direction. The resulting beam sizes are listed in Table~\ref{tab:table1}. Following the order of the source names in the table, the root mean square (RMS) noise levels in the line-free channels of the NH$_3$(1,1) line are approximately 0.4, 0.2, 0.5, and 0.2~K per channel of 0.3~km~s$^{-1}$, respectively.

\begin{figure*}[ht]
  \centering

    \includegraphics[width=0.49\textwidth]{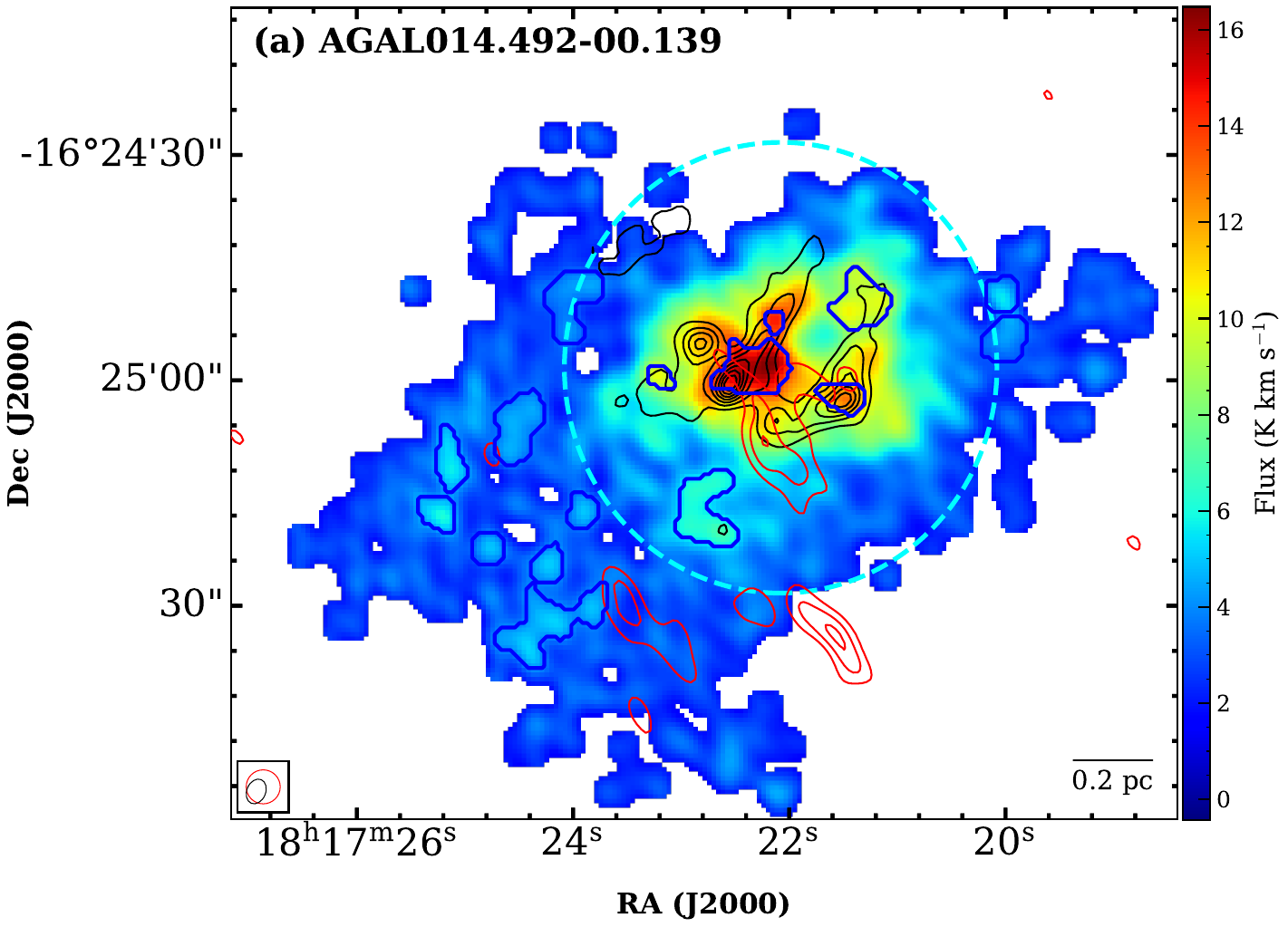}
    \includegraphics[width=0.43\textwidth]{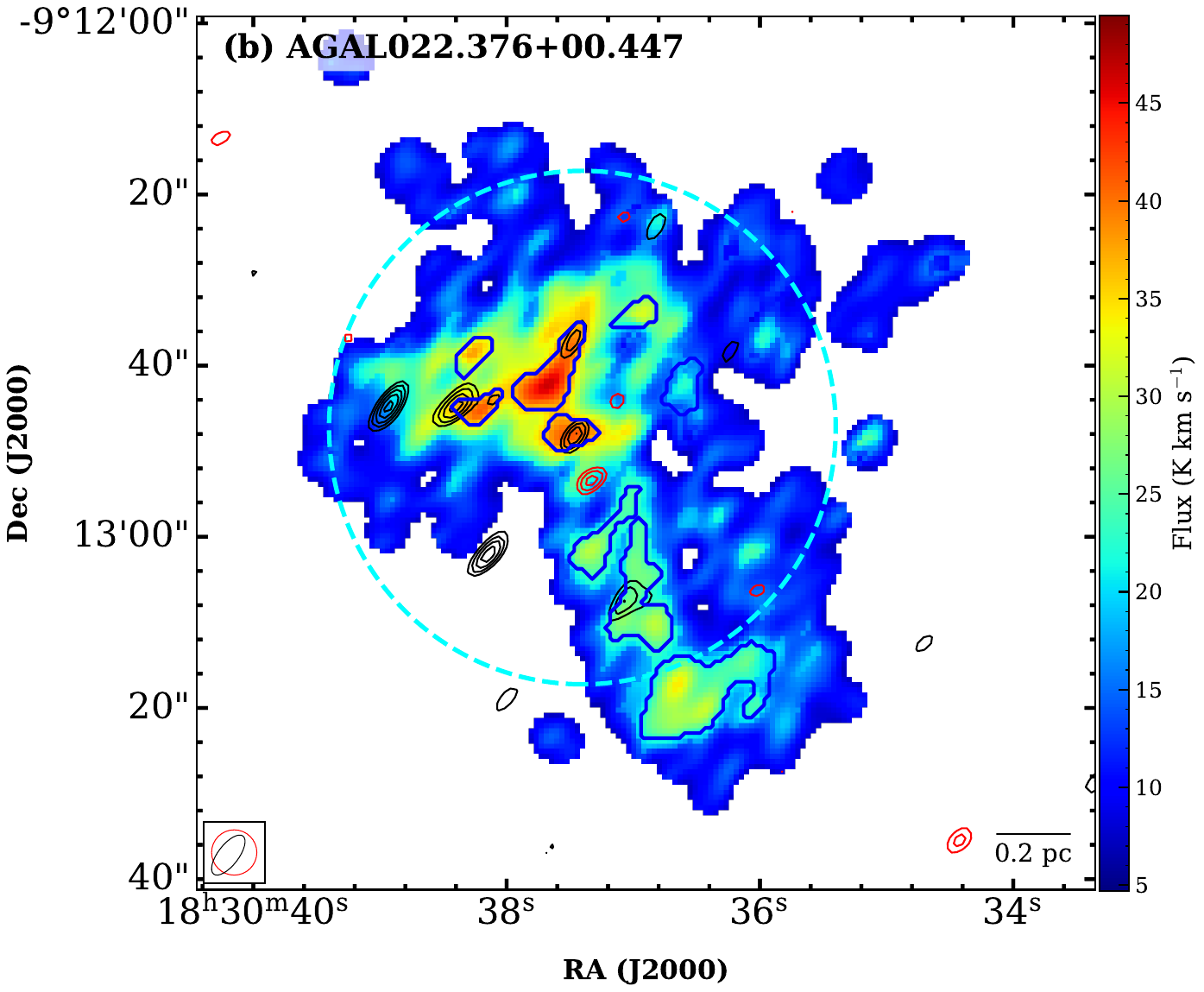}
    \includegraphics[width=0.49\textwidth]{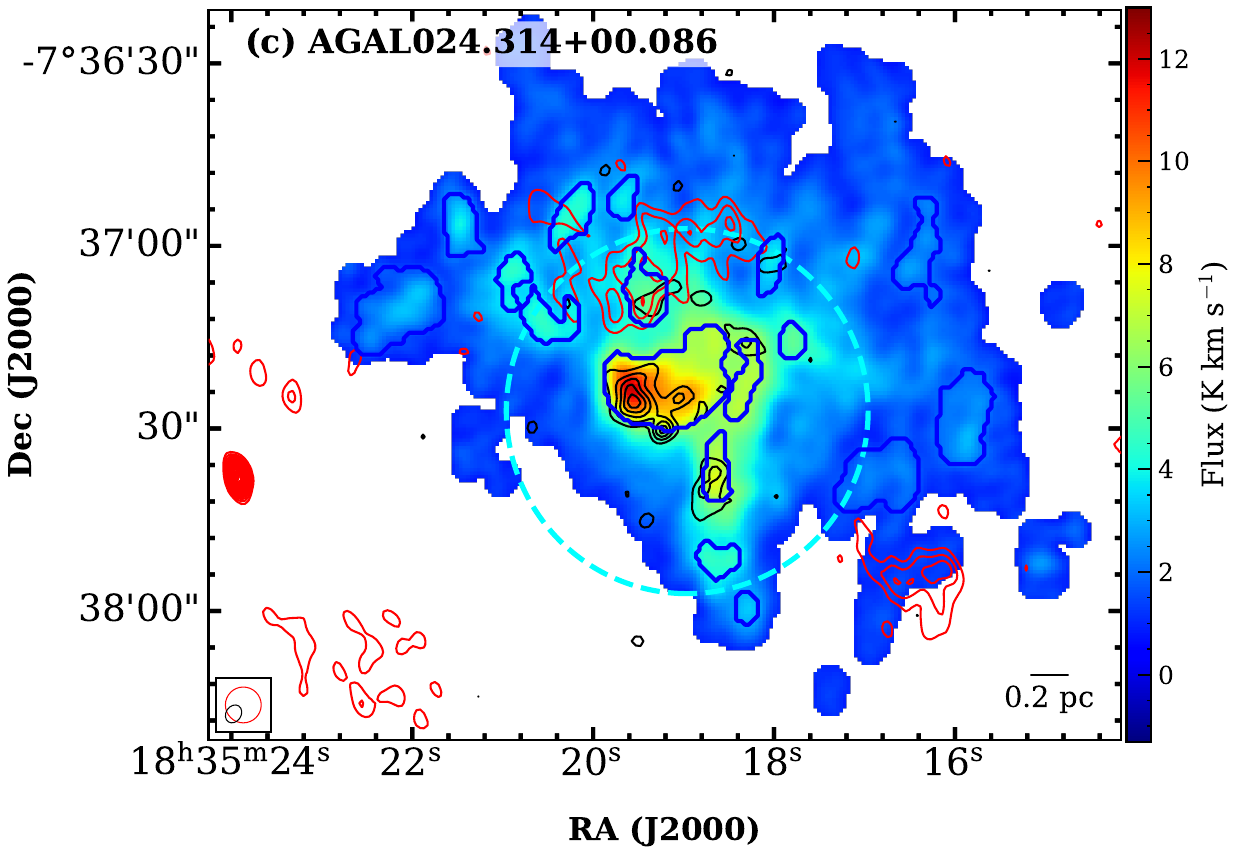}
    \includegraphics[width=0.43\textwidth]{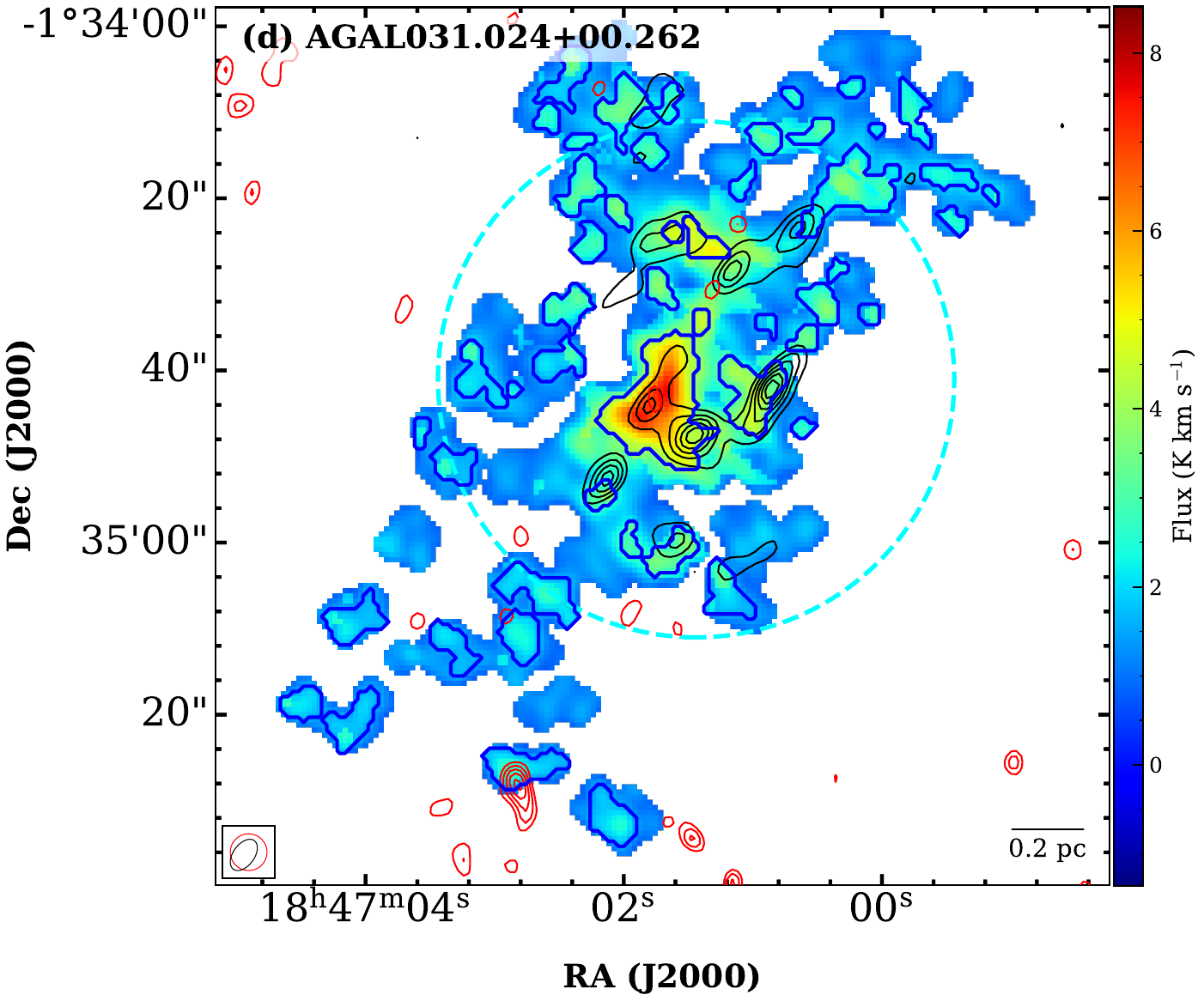}
    \caption{The background shows the velocity-integrated intensity of the NH$_3$(1,1) inner left satellite line, overlaid with black contours of the 1.3~mm continuum and red contours of the 1.3~cm continuum, except for panel~(d), which shows the inner left satellite line of the blue component and the inner right satellite line of red components for AGAL031.024+00.262. The noise level ($\sigma$) of the velocity-integrated intensities are 0.6, 1.1, 0.3, and 0.3~K~km~s$^{-1}$, respectively.
 The blue contours show the leaf structures identified by the astrodendro framework. The dotted circle marks the field of view in the SMA observation. The beam size in the left bottom consists of the ammonia lines for the red ellipse and 1.3 mm continuum for the black ellipse. The adopted contours are different for each of the four sources. The root mean square (RMS) noise levels are listed in Table~\ref{tab:table2}. 
(a) The contours start at $3\cdot\mathrm{RMS}$, stepping with $4\cdot\mathrm{RMS}$ for the black contours and $2\cdot\mathrm{RMS}$ for the red contours. 
(b) The contours start at $3.5\cdot\mathrm{RMS}$, stepping with  $1\cdot\mathrm{RMS}$ for both black and red contours. 
(c) The contours start at $3\cdot\mathrm{RMS}$,  stepping with $4\cdot\mathrm{RMS}$ for the black contours and $2\cdot\mathrm{RMS}$ for the red contours. 
(d) The contours start at $3\cdot\mathrm{RMS}$, stepping with $3\cdot\mathrm{RMS}$ for the black contours and $1\cdot\mathrm{RMS}$ for the red contours. 
}
\label{fig:fig1}
\end{figure*}
\section{Methods}
\label{sec:met}

The \amm\ (1,1) and \amm\ (2,2) inversion transitions were used to derive the kinematic properties of the four clumps. The field of view of the VLA observations is about three times larger than previous observations of these clumps from the SMA \citep{2019ApJ...886..130L}, allowing us to trace more extended structures. We employed the combined VLA and GBT data for the subsequent analysis.

\subsection{Extracted structures from the NH$_3$(1,1) inner satellite group}

Dense gas structures can be traced by both dust continuum emission and ammonia line emission.
Before applying optical-depth corrections, the integrated intensity distribution of the NH$_3$(1,1) main components, which are generally optically thick, does not reliably reflect the underlying gas density structure (see Sect.~\ref{sect:tau} for a detailed discussion). To compare the physical properties of dense cores identified from the continuum and ammonia data, we extracted dense cores from the  intensity map of NH$_3$(1,1) inner satellite line using the \texttt{dendrogram} algorithm \citep{2008ApJ...679.1338R}, as shown in Fig~\ref{fig:fig1}.

In the \texttt{astrodendro} framework\footnote{http://www.dendrograms.org}, structures are hierarchically identified as `leaves`, `branches`, and `trunks`, corresponding to substructures, intermediate structures, and the overall emission, respectively. A leaf represents an individual structure that satisfies a minimum intensity threshold, a peak intensity criterion, and a minimum difference from $\sigma$, where $\sigma$ is the noise level of the velocity-integrated intensity, calculated as $\mathrm{RMS}\cdot\sqrt{\Delta V\cdot\delta V}$. Here, $\Delta V$ is the velocity integration range, and $\delta V$ is the velocity resolution. We integrated the \amm(1,1) emission over velocity ranges of 43.7--52.1 \kms, 73.1--81.2 \kms, and 103.2--111.3 \kms\ for AGAL014.492--00.139, AGAL022.376+00.447, and AGAL024.314+00.086, respectively. For AGAL031.024+00.262, the two integration intervals used were 67.5--73.2 \kms\ and 101.7--106.8 \kms. These intervals target the left inner satellite of the blue-shifted component and the right inner satellite of the red-shifted component. This ensures that the contributions from different velocity features and their respective hyperfine satellite components are properly captured. In this work, we adopt a minimum intensity threshold of $5\sigma$, a peak intensity threshold of $7\sigma$, and a minimum $\Delta\sigma$ of $1.5\sigma$. In addition, we mask out leaves smaller than two-thirds of the synthesized beam size to eliminate spurious detections. 

\subsection{Deriving physical parameters from line fitting}
\label{sec:3.2.2}

We used the \texttt{pyspeckit} package to simultaneously fit the \amm(1,1) and \amm(2,2) spectra  with the \texttt{cold\_ammonia} model \citep{2008ApJS..175..509R, 2009ApJ...697.1457F, 2015PASP..127..266M}, implemented within the integrated development environment \citep{2011ascl.soft09001G, 2022AJ....163..291G}. The fitting assumes a common excitation temperature for all hyperfine components and adopts the Rayleigh-Jeans approximation. The model parameters with the initial guesses are excitation temperature (T$_{ex}$=5 K), para-ammonia column density ($\log_{10}(N_{tot})=15$ cm$^{-2}$ ), kinetic temperature (T$_{k}$=15 K), velocity dispersion ($\sigma_{v}$) from the second-moment map of the \amm(1,1) main line, and the velocity of the local standard of rest (V$_{LSR}$) from the first-moment map of the \amm(1,1) main line. The best-fit parameters are obtained using the MPFIT algorithm, which performs a non-linear least-squares minimization  \citep{2009ASPC..411..251M}.

Parameter maps of AGAL031.024+00.262 were generated using the extended implementation of \texttt{pyspeckit}\footnote{\url{https://github.com/autocorr/nestfit}}, which employs the same \texttt{ammonia\_model} but compiled in \texttt{Cpython} for improved computational efficiency. To discriminate between one and two velocity components, which up to a maximum separation of $\sim$18~km\,s$^{-1}$, we used Bayesian evidence ($Z$) computed via nested sampling \citep{2004AIPC..735..395S}. Model comparisons were performed using the Bayes factor $K^{i}_{j}=Z_{i}/Z_{j}$, where $K^{i}_{j}>11$ indicates that model $i$ is strongly preferred \citep{2020ApJ...892L..32S}. Specifically, $K^{1}_{0}$ distinguishes single-component spectra from non-detections, and $K^{2}_{1}$ identifies spectra that require two components.

For AGAL014.492–00.139 and AGAL024.314+00.086, the typical velocity separation between components is $\sim$3 km\,s$^{-1}$. We find that most ammonia cores cannot be reliably decomposed into two components at the current angular resolution, except in regions where the velocity separation exceeds 3 km\,s$^{-1}$. The five hyperfine components of \amm(1,1) are heavily blended in two-component regions, indicating the complex kinematic environment of IRDCs. Therefore, the spectra within the regions of each dense core are spatially averaged for subsequent analysis.

The velocity dispersion ($\sigma_{v}$) consists of both thermal and nonthermal components, as well as the contribution from the spectral channel width ($\sigma_{\rm chan}$). Therefore, the nonthermal component is expressed as:
\begin{equation}
    \sigma_{non}^{2} = \sigma_{v}^{2} - \sigma_{\rm th,NH_3}^{2} - \sigma_{\rm chan}^{2}.
\end{equation}
The thermal component of ammonia is given by
$\sigma_{\rm th,NH_3} = \sqrt{k_{B}T / \mu_{\rm NH_3}}$,
where $\mu_{\rm NH_3} \approx 17$~amu. The sound speed is
$c_{\rm s} \approx 0.22$~km\,s$^{-1}$ at T$_{k}=15$~K for a mean molecular weight $\mu = 2.37$.  
The one-dimensional Mach number is then defined as
$\mathcal{M} = \sigma_{non}/c_{\rm s}$,
where $\mathcal{M} < 1$, $\mathcal{M}=$1-2, and $\mathcal{M} > 2$ correspond to subsonic, transonic, and supersonic motions, respectively \citep{2018ApJ...861...77M}.  
The intrinsic linewidth with smaller than the channel width of $\sigma_{\rm chan}=0.12$~km\,s$^{-1}$ is not considered in this study. Total parameters are list in Table \ref{tab:tab3}.

\begin{table*}[]
\centering
\caption{Comparison of the 1.3~cm and 1.3~mm continuum emission. }
\begin{tabular}{cccccccccc}
\hline
\hline
Source  & 1.3 cm & S/N$_{cm}$ & RMS$_{cm}$        & 1.3 mm & RMS$_{mm}$        & S/N$_{mm}$ & S$_{\nu}$& L$_{T}$ & N$_{uv}$ \\
        & 10$^{-3}$K  &     & 10$^{-3}$K & 10$^{-2}$K & 10$^{-3}$K & &10$^{-5}$Jy & 10$^{-10}$ W Hz$^{-1}$ & 10$^{44}$   s$^{-1}$               \\ \hline
AGAL014.492–00.139 & 16.2        & 7.5 & 2.1        & 41.1      & 2.2        &186    &14.4 &25.7 & 2.5        \\
AGAL022.376+00.447 & 4.5        & 6.4 & 3.5       & 11.5      & 1.1        & 105     &1.9& 5.4  &0.5     \\
AGAL024.314+00.086 & 19.6       & 8.5 & 2.3        & 49.5       & 1.3        & 381   &11.4 & 5.4 & 2.9          \\
AGAL031.024+00.262 & 9.9         & 7.5 & 1.3       & 25.1       & 0.62       & 404    &4.0 & 30.8 &1.9         \\ \hline
\end{tabular}
\tablefoot{The first column lists the source names. Columns 2–4 present the peak intensity, S/N, and RMS of the 1.3~cm continuum, respectively. Columns 5–7 show the corresponding peak values for the 1.3~mm continuum and the theoretical S/N. Columns 8–10 provide the flux density, thermal spectral luminosity, and the production rate of Lyman continuum photons.}
\label{tab:table2}
\end{table*}

\section{Result}
\label{sec:result}
\subsection{Continuum emission}
\label{sec:3.1}

The continuum emission can be used to trace the evolutionary stages of star formation \citep[e.g.,][]{2001ApJ...549..979K, 2019MNRAS.485.2895E}. 
Based on the 1.3\,cm continuum observations, 
the peak brightness temperatures range from 4.5\,mK in AGAL022.376+00.447 to 19.6\,mK in AGAL024.314+00.086, 
except for a compact point source in AGAL024.314+00.086 that reaches a peak brightness temperature of $\sim$ 0.88\,K. 
Extended continuum structures (0.4$\sim$0.7 pc) are detected in AGAL024.314+00.086 and AGAL014.492–00.139, 
while the emission regions in the other sources are generally comparable to the beam size.

To confirm whether the 1.3 cm continuum originates from dust thermal emission, we compare its intensity distribution with that of the 1.3 mm continuum. As shown in Fig.~\ref{fig:fig1}, the 1.3 cm continuum emission is offset from that of the 1.3 mm continuum emission obtained from SMA observations \citep{2019ApJ...886..130L}.
The 1.3\,mm continuum emission is dominated by thermal dust radiation. Assuming the optically thin limit following equation 2 of \cite{2012MNRAS.425.3094C} and adopting a spectral index of $\beta \sim 1.5$, the flux density scales as:
\begin{equation}
    S_{\nu} \propto \nu^{\beta} B_{\nu}(T) \propto \frac{\nu^{4.5}}{\mathrm{e}^{h\nu/k_{B}T} - 1},
\end{equation}
where $S_{\nu}$ is the flux density at frequency $\nu$, $B_{\nu}(T)$ is the Planck function, $T$ is the dust temperature, respectively. Based on this relation, the 1.3\,mm intensity can be extrapolated to the 1.3\,cm continuum  to further confirm the evolutionary stage, as listed in Table \ref{tab:table2}.
The brightness temperature derived from the spectral index of 4.5 is significantly lower at 1.3 cm compared to 1.3 mm, consistent with thermal dust emission decreasing rapidly with decreasing frequency. However, the observed 1.3 cm emission is unlikely to be thermal dust radiation, as the extrapolated flux would be below the detection sensitivity of the SMA observations in \cite{2019ApJ...886..130L}. 

Free--free emission, which typically originates from late evolutionary stages of star formation, generally exhibits brightness temperatures much greater than 1~K at 1.3~cm continuum, exceeding the maximum value of 0.88~K found in our results. In addition, we calculate the production rate $N_{\rm uv}$ of Lyman continuum photons to further assess the possibility of free--free emission given the detection sensitivity of the VLA observations. The production rate $N_{\rm uv}$ is related to the thermal spectral luminosity $L_T$ and characterizes HII regions photoionized by hot stars \citep{1992ARA&A..30..575C}. The value of $N_{\rm uv}$ can be derived using equation~2 in \cite{1992ARA&A..30..575C}:
\begin{equation}
\left( \frac{N_{\rm uv}}{\rm s^{-1}} \right) \geq 6.3 \times 10^{52}
\left( \frac{T_e}{10^4\,{\rm K}} \right)^{-0.45}
\left( \frac{\nu}{\rm GHz} \right)^{0.1}
\left( \frac{L_T}{10^{20}\,{\rm W\,Hz^{-1}}} \right),
\end{equation}
where the electron temperature is $T_e = 8000$~K, the observing frequency is $\nu = 23.4$~GHz, and $L_T = 4\pi D^2 S_\nu$.
This equation assumes optically thin emission and therefore provides a limit on $N_{\rm uv}$. For the four clumps, we derive the flux density $S_\nu$ within a circular region with a radius equal to the beam major axis, centered on the peak position. The compact point source in AGAL024.314$+$00.086 is excluded from the calculation. The resulting parameters are listed in Table~\ref{tab:table2}.

We find that the values of $N_{\rm uv} = 5\times10^{43} \sim 2.5\times10^{44}\ {\rm s^{-1}}$
for the four clumps corresponds to spectral type of B2--B3,
assuming a single zero-age main-sequence star as the ionizing source,
with stellar mass of 7$\sim$8~$M_\odot$ \citep{1973AJ.....78..929P, 1996ApJ...460..914V, 1997ApJ...489..698H,
1998ApJ...501..192D, 2005A&A...436.1049M}.
The 1.3~cm continuum emission is possibly produced by shock-induced ionization from outflows or jets and is not associated with HII regions \citep{2024A&A...682A.180D}. Therefore, we conclude that the seventeen dense cores in the four clumps reported by \cite{2019ApJ...886..130L} that show no outflow are likely in the early stages of star formation.

\subsection{ The properties of ammonia cores}

We initially identified 92 dense cores from the velocity-integrated intensity of NH$_3$(1,1) inner satellite line , which is generally optically thin. Among them, 61 cores satisfy the size criterion of being larger than $2/3$ of the beam size, which removes spurious detections and allows them to be classified as ammonia cores. Within this sample, 38 cores exhibit a single velocity component, while 23 show multiple velocity components in the NH$_3$(1,1) main line.

\begin{figure*}
    \centering
    \includegraphics[width=1\linewidth]{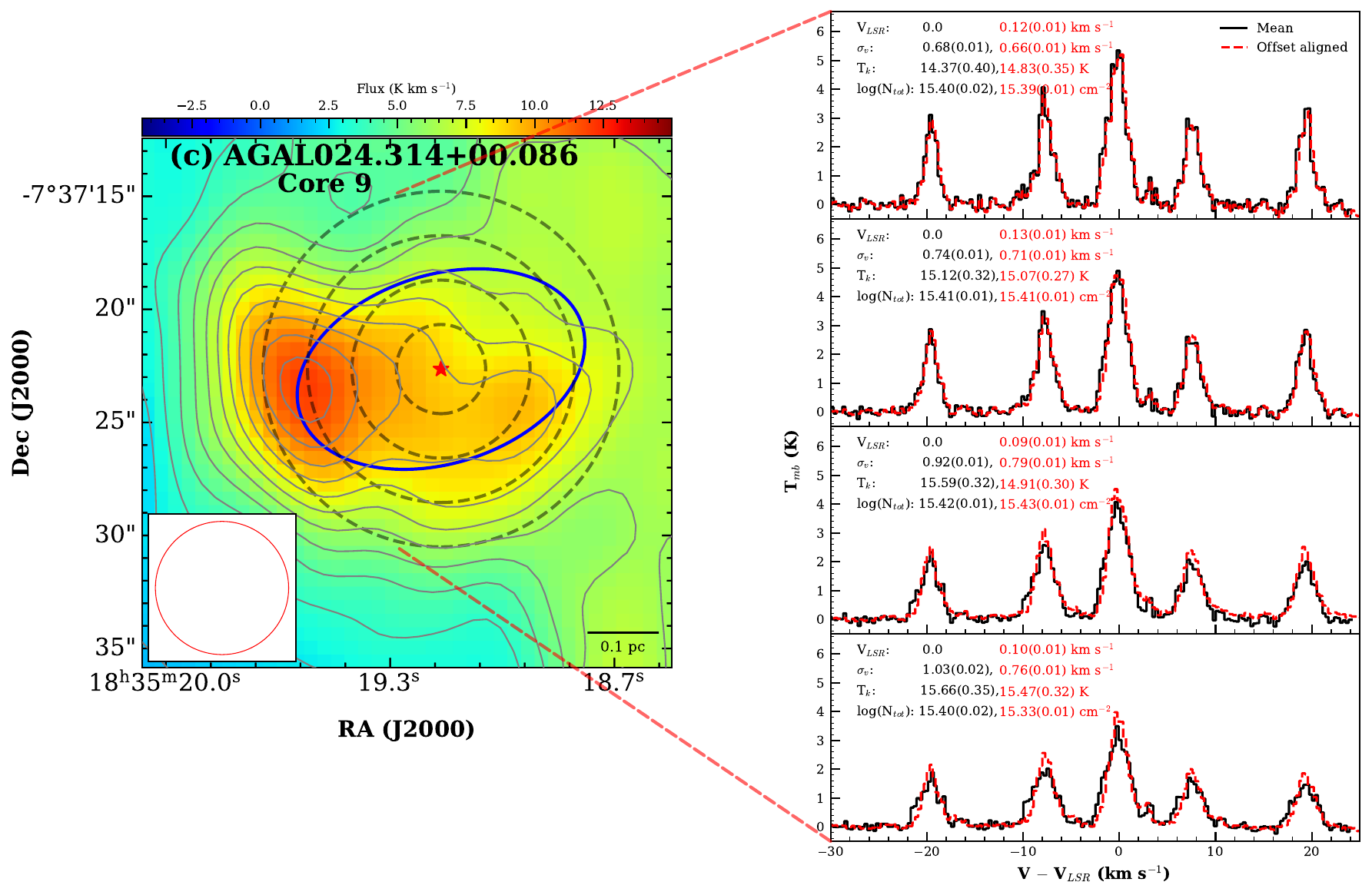}
\caption{
The left panel shows example of identified dense core, outlined by a blue ellipse with its center marked by a red star. The colored background is the NH$_3$(1,1) satellite-line intensity, also shown with gray contours. The red circle indicates the beam sizes of the \amm(1,1) line. The right panel shows the averaged NH$_3$(1,1) line profile for each annular region as a black solid line, and the averaged spectral profile with the pixel-by-pixel velocity axis referenced to $V_{\rm LSR}$ as a red dashed line. The fitting parameters are denoted by black and red text, respectively. The $V_{\rm LSR}$ values of the spectra are 114.26, 114.27, 114.18, and 114.07 \kms\ for the black lines before they are aligned to zero. From top to bottom, the spectra correspond to regions from the core center to the outer areas.
}
\label{fig:fig2}
\end{figure*}

\subsubsection{Averaged physical parameters}
For each of the 61 ammonia cores, the NH$_3$(1,1) and NH$_3$(2,2) spectra were averaged over regions corresponding to the core sizes due to S/N limitations . The averaged spectra were then fitted, and the derived physical parameters are summarized in Table~\ref{tab:tab3} of appendix.
The kinetic temperatures span from 10.53~K to 26.61~K. The ammonia column densities range from $3.07\times 10^{14}$~cm$^{-2}$ to $8.19\times 10^{15}$~cm$^{-2}$. The velocity dispersions vary between 0.18~km\,s$^{-1}$ and 1.84~km\,s$^{-1}$. The observed line profiles are affected not only by optical depth effects but also by local protostellar feedback such as molecular outflows. Twelve ammonia cores have counterpart 1.3~mm continuum emission, among which five are associated with CO\,$J=2$-1 outflows \citep{2019ApJ...886..130L}.

The ammonia cores showing single velocity component, the thermal component ranges from 0.20 to 0.24~km\,s$^{-1}$, while the nonthermal component spans from 0.22 to 1.2~km\,s$^{-1}$. For AGAL031.024+00.262, we adopt a uniform kinetic temperature of T$_{k}$=13~K for ammonia cores whose the NH$_3$(2,2) main line intensity have S/N $<$ 3. All ammonia cores in AGAL014.492–00.139, AGAL022.376+00.447, and AGAL024.314+00.086 show supersonic motions, whereas four ammonia cores in AGAL031.024+00.262 exhibit sonic levels.

Nine cores show multiple velocity components, each consisting of a narrow and a broad feature. For example, Core~9 of AGAL014.492–00.139 exhibits a sonic component with $\mathcal{M}=1.18$ and a supersonic component with $\mathcal{M}=4.06$. Core~8 of AGAL024.314+00.086 similarly shows a supersonic component ($\mathcal{M}=5.47$) accompanied by a nearly sonic one ($\mathcal{M}=1.01$).
The kinetic temperatures of the narrow and broad components differ by $>$4~K. Thermal motions dominate the narrow component in Core 8, whereas turbulence drives the broad component. This contrasts with other ammonia studies, where both components are typically supersonic \citep[e.g.,][]{2020A&A...640L...6C}. The thermal to nonthermal transition may indicate progressive turbulent energy dissipation \citep{2014ApJ...789...83T}.

AGAL031.024+00.262 contains two molecular clumps with distinct physical conditions. The blue-shifted components are systematically colder by $\sim2$~K and have ammonia column densities approximately a factor of two lower than those of the red-shifted components. Two cores in this region show both subsonic and supersonic components. In Core~16, the blue component is supersonic ($\mathcal{M}=1.8$), whereas the red component is sonic ($\mathcal{M}=0.43$). Conversely, in Core~24, the blue component is subsonic ($\mathcal{M}=0.58$) and the red component is strongly supersonic ($\mathcal{M}=2.89$). The dominance of thermal motions in some of these components indicates that turbulence has largely dissipated in the central regions, resulting in quiescent gas characteristic of very early evolutionary stages of star formation.

\subsubsection{Internal radial gradient of \nonV}
\label{sec:4.2.2}
The multiple velocity components in the NH$_3$(1,1) spectra cause line broadening that cannot be cleanly decomposed on pixel-by-pixel basis when the velocity separation is smaller than $\sim 3$~km s$^{-1}$. As a result, reliable pixel-by-pixel maps of velocity dispersion and kinetic temperature cannot be obtained for all dense cores. To investigate the radial variations of kinematic and thermal properties, we averaged the spectra based on a series of concentric annuli. From the outer envelope toward the core center, the annuli were spaced at intervals of one-third of the major axis of beam. The maximum radius was defined as the sum of the major axis of ammonia core and half of the beam size.Within each annulus, the spectra were spatially averaged to improve the S/N. 
The annular regions may contain velocity gradient  or vary systemic velocity across different positions, which could broaden the linewidth and introduce a bias in the radial gradient of \nonV. 
We used the method described by \cite{1993ApJ...406..528G} to fit the velocity field of the dense cores. The typical velocity gradient ranges from 0.5 to 3.8 \kms\ pc$^{-1}$. 
Although the current spatial resolution does not allow us to disentangle the contributions of infall and/or rotation to the linewidth. 
The influence of velocity gradients and variations on the linewidth can be significantly mitigated by aligning each spectrum to a common velocity center (here, V$_{LSR,\ \mathrm{com}} = 0$) before averaging them over the annular region (see Fig.~\ref{fig:fig2}).

The averaged NH$_3$(1,1) and NH$_3$(2,2) spectra in each subregion were then fitted simultaneously using the \texttt{cold\_ammonia} model implemented in \texttt{pyspeckit}, yielding radial profiles of velocity dispersion and kinetic temperature. A sample of the spectral line is shown in Fig.~\ref{fig:fig2}, where we compare the direct average of spectra in the annular region to those with aligned spectra. The spectral line shows systematically lower linewidths after the velocity alignment, effectively reducing the influence of variations in the systemic velocity across different positions within the annulus. Therefore, we apply the result derived from the velocity alignment spectra to our work. The radial gradient of \nonV\  for AGAL031.024+00.26 is presented in Fig.~\ref{fig:fig3}. The other source present in  Fig. \ref{fig:figB} of appendix. The derived parameters are listed in Table~\ref{tab:tab4}.

As shown in Fig.~\ref{fig:fig3} and Fig. \ref{fig:figB}, the radial behavior of \nonV\ varies significantly among the ammonia cores. The \nonV\ increases outward, exhibiting supersonic motions on scales of 0.01$\sim$0.11~pc toward the five cores in AGAL022.376+00.447 and AGAL024.314+00.086. In addition to these ammonia cores, Core~1 of AGAL014.492–00.139 and Core~11 of AGAL024.314+00.086 span a smaller radial range, showing only minor variations of 0.01~km~s$^{-1}$ and 0.02~km~s$^{-1}$, which are not included in the following discussion. Core~15 of AGAL024.314+00.086, on the other hand, displays a much larger change of 0.24~km~s$^{-1}$ across a distance of 0.0157$\sim$0.1091~pc. These trends indicate that the turbulence intensity generally decreases toward the core centers. If such turbulence dissipation causes the nonthermal component to approach sonic or subsonic levels, the cores may experience faster gravitational collapse than those dominated by supersonic turbulence. For example, Core~1 of AGAL031.024+00.262 shows a \nonV\ of 0.23~km\,s$^{-1}$ at its center of $\sim$0.0086~pc, transitioning to a transonic value of 0.39~km\,s$^{-1}$ in the outer region of $\sim$0.0423~pc.
In contrast, Cores~6, 8, and 9 in AGAL031.024+00.262 exhibit enhanced turbulence toward their centers, with $\Delta$\nonV\ from 0.14 to 0.20~km s$^{-1}$.

The ammonia cores in AGAL031.024+00.262  associated with multiple velocity components reveal further kinematic complexity. In Cores~17 and~18, which coincide with the 1.3~mm continuum emission, the blue-shifted components show a pronounced outward increase in \nonV\ in contrast to the red-shifted components.
The difference in \nonV\ between the distinct velocity components ranges from 0.02 to 0.15~km~s$^{-1}$. In Core~21, the spatial variation of \nonV\ is 0.23~km~s$^{-1}$ for the blue component and 0.14~km~s$^{-1}$ for the red component. Notably, both components trace a transition from the sonic to the supersonic regime. In Core~22, a similar decreasing trend is observed in both components, with differences of 0.05~km~s$^{-1}$ in the blue component and 0.12~km~s$^{-1}$ in the red component.

The kinematic temperature profiles across the annuli remain relatively flat within individual cores. Central temperatures range from 10.4 to 24.8~K, while outer regions spans from  11.0 to 16.1~K. As shown in Fig.~\ref{fig:pltpara}, the kinematic behaviors in ammonia cores exhibit a similar distribution. The \nonV\ increases with T$_k$, indicating that feedback from ongoing star formation activities enhances the stability of dense cores and highlights the significance of turbulence in differentiating evolutionary stages.

\section{Discussion}
\label{sec:disc}
\subsection{Turbulence dissipation and enhancement}
\label{sect:turb}
Stable dense cores require sufficient internal support through turbulence, magnetic pressure, and thermal motions to counteract self-gravity \citep{2007ARA&A..45..565M}. As described above, both thermal and nonthermal motions are present in the ammonia cores under varying local conditions, influenced by internal and external feedback processes. \citet{2022MNRAS.517..885O} proposed three evolutionary phases characterized by different turbulence regimes dependent on spatial scale and gas density. Using magnetohydrodynamic simulations, they modeled three-dimensional (3D) velocity dispersion gradients in 0.1~pc cores and suggested a typical core lifetime of order $10^{5}$~yr.

\begin{figure*}[!t]
    \centering
        \includegraphics[width=1\linewidth]{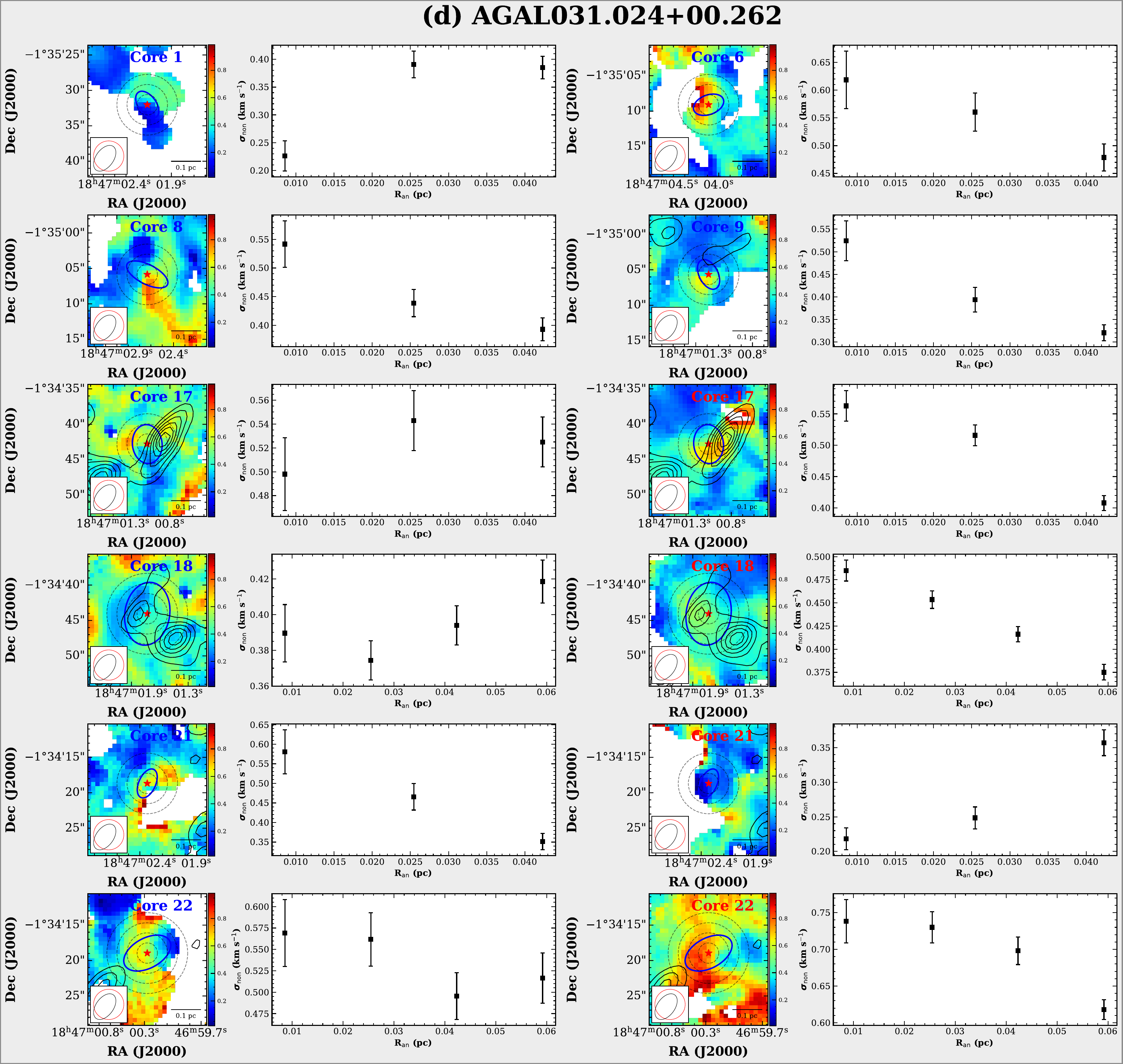}\\[1ex]
    \vspace{1em}
    \caption{The colored background in the left panel denotes the \nonV\ map of the NH$_{3}$(1,1) main line. The dashed circular ring outlines the averaged regions used for the NH$_{3}$(1,1) and NH$_{3}$(2,2) spectra. The black contours show the 1.3 mm continuum with the same contour levels as those in Fig. \ref{fig:fig1}.  The black circle indicates the beam size of the 1.3 mm continuum observation, while the red circle denotes the beam size of the NH$_{3}$(1,1) line. The right panel displays the radial variation of \nonV\ from the center to the outer regions, marked in black. For the source AGAL031.024+00.262, the two colored labels represent the blue- and red-shifted components of the NH$_{3}$(1,1) main line.}
    
    \label{fig:fig3}
\end{figure*}

\begin{figure*}[t]
    \centering
    \resizebox{\hsize}{!}{\includegraphics[width=0.5\linewidth]{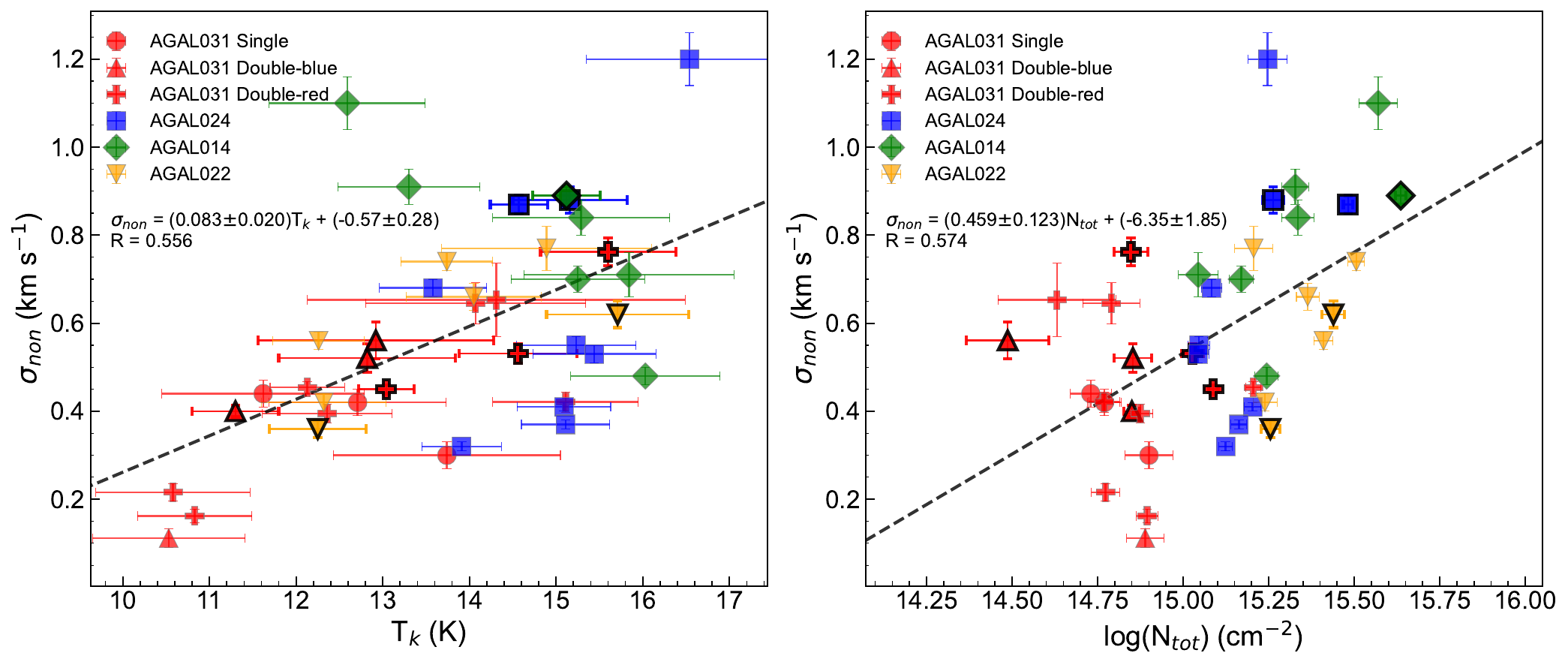}}
    \caption{The figure shows the correlations between \nonV\ and both T$_{k}$ and N$_{tot}$ for the ammonia cores with single velocity components, as well as for the blue- and red-shifted components of AGAL031.024+00.262. Different colored markers represent the averaged parameters of each core, and the black outlines indicate the ammonia cores that are associated with the 1.3\,cm continuum emission.}
    \label{fig:pltpara}
\end{figure*}

Comparing our observational results with their predictions, we find that five ammonia cores, out of a total of 33 cores in AGAL014.492–00.139, AGAL022.376+00.447, and AGAL024.314+00.086, exhibit similar behavior on a characteristic scale of $\sim$0.1~pc, in which the nonthermal velocity component decreases toward the core center.
 The minimum observed \nonV\ of 0.2~km\,s$^{-1}$ lies at the lower end of the 3D velocity dispersion gradients modeled by \citet{2022MNRAS.517..885O}. However, since NH$_3$ linewidths trace only the one-dimensional projection, the true three-dimensional velocity dispersions must be larger. This implies that turbulence remains significant and cannot be ignored, both before and after the collapse of dense cores.
In these five ammonia cores, the central regions exhibit reduced \nonV, providing less nonthermal motions support against gravity compared to the outer envelopes. This trend suggests that turbulence dissipates toward higher-density regions, thereby facilitating inward gas accretion in the absence of additional forces opposing gravitational collapse \citep{2025ApJ...988...82M}.

In AGAL031.024+00.262, the radial behavior of \nonV\ is more diverse. One core shows an increase in \nonV\ within 0.0423~pc, while three single velocity component cores show decreasing trends from the center to the outer region. For the multiple velocity component cases (Cores~17, 18, and~21), \nonV\ behaves differently between velocity components that Cores~17 and~18 show enhanced \nonV\ in the blue-shifted components but decreasing values in the red-shifted ones. The revealed CO outflow in Core~17 indicates the presence of at least one embedded protostar, suggesting that the central turbulence enhancement is likely driven by feedback from ongoing star formation. The outflow likely originates from the red-shifted component, while the blue component shows an outward increase in \nonV, implying that star formation may be occurring on smaller scales ($<$0.01~pc). In contrast, the decrease in \nonV\ for both the blue-shifted and red-shifted components in Core~22 suggests that ongoing star formation activities are affecting the surrounding environment, despite the absence of detectable 1.3~mm continuum emission and outflow in this study.

Compared with the coherent cores identified by \cite{2021A&A...648A.114C} in L1688, the velocity dispersion increases from the core center (0.2$\sim$0.4 \kms) to the shell ($\sim$0.6 \kms) surrounding the core, and both exhibit two velocity components with a similar trend. In addition, they confirm that the velocity dispersion difference $\Delta\sigma_{v}$ is 0.15$\sim$0.25 \kms\ for a transition scale of 0.2 pc. In this work, Core~1, and Core~21, the turbulence has largely dissipated on small scales ($\sim$0.1 pc), leaving the gas predominantly thermally supported. Thermal pressure dominates in the central regions of the blue component of Core~1 and the red component of Core~21. These components exhibit low kinetic temperatures (e.g., T$_{k} = 10.5$~K for Core~21) and low total column densities (N$_{tot} = 5.9 \times 10^{14}$~cm$^{-2}$), compared with typical values of $\sim$14~K and $10^{15}$~cm$^{-2}$ for the other cores. These cores appear dynamically quiescent and may soon undergo gravitational collapse to form prestellar or protostellar objects, consistent with the second phase described by \citet{2022MNRAS.517..885O}. Ammonia cores that exhibit outward-increasing \nonV\ profiles are likely at earlier evolutionary stages.

Due to the low S/N ($<$3) of the NH$_3$(2,2) main line  and the presence of multiple velocity components across three clumps, reliable radial density profiles cannot be obtained to further constrain the evolutionary states. Higher angular resolution and higher sensitivity observations at sub-parsec scales will be essential to reveal the transitions between thermally dominated and turbulence-dominated regions within these ammonia cores.

\subsection{Evolutionary stage of the ammonia cores}
\label{sect:feedback}
As described in Sect.~\ref{sec:3.1}, the 1.3~mm and 1.3~cm continuum emission show positional offsets within the SMA field of view. The 1.3~mm continuum emission may primarily trace thermal dust emission in cold regions with temperatures of $\sim$15~K for this work, which decreases with frequency following a spectral index of 4.5 \citep{2012MNRAS.425.3094C}. The flux density therefore drops by nearly four orders of magnitude from 1.3~mm to 1.3~cm, requiring extremely high sensitivity at longer wavelengths that is not achieved in the current VLA data. The 1.3~cm continuum emission exhibits extended structures in AGAL014.492–00.139 and AGAL024.314+00.086, which likely trace nonthermal dust emission from background sources and are not further investigated in this work.

Ammonia is also used to trace dense gas and to probe the evolutionary stages of dense cores \citep{2013MNRAS.432.3288S}.  We identify twelve ammonia cores associated with 1.3~cm continuum emission and five cores with detected outflows. Starless cores that are transonic or supersonic and show no detected outflows, yet are located near protostars, are likely perturbed by these outflows and therefore exhibit broader linewidths \citep{2013MNRAS.432.3288S}.
From the CO~$J=2\!-\!1$ results of \citet{2019ApJ...886..130L}, the outflow lengths typically exceed 0.1~pc, comparable to the sizes of the ammonia and 1.3~mm continuum cores. The interior feedback can result in the heated component of molecular gas and inject turbulence. The correlations of \nonV\  present middle trend ($R\sim$0.56) comparing with the  T$_{k}$ and N$_{tot}$, as shown in Fig \ref{fig:pltpara}. The ammonia cores associated with the 1.3~mm continuum do not show significant differences in T$_{k}$ or N$_{tot}$ among the four clumps, indicating that the dense gas traced by both ammonia and dust shows a similar physical condition.

\begin{figure*}[t]
    \centering
    \includegraphics[width=0.34\linewidth]{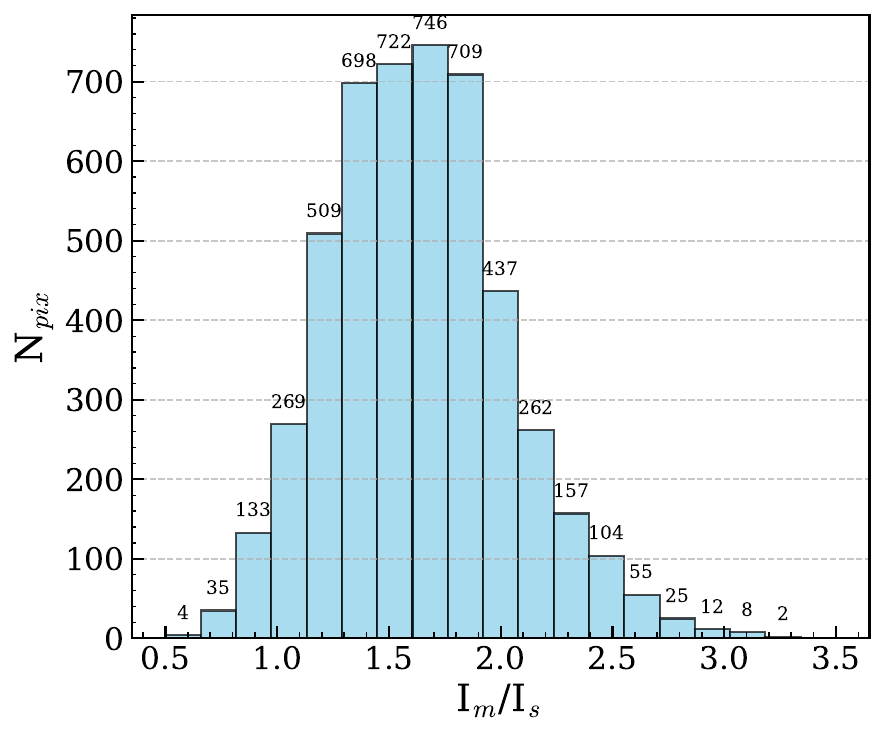}\includegraphics[width=0.34\linewidth]{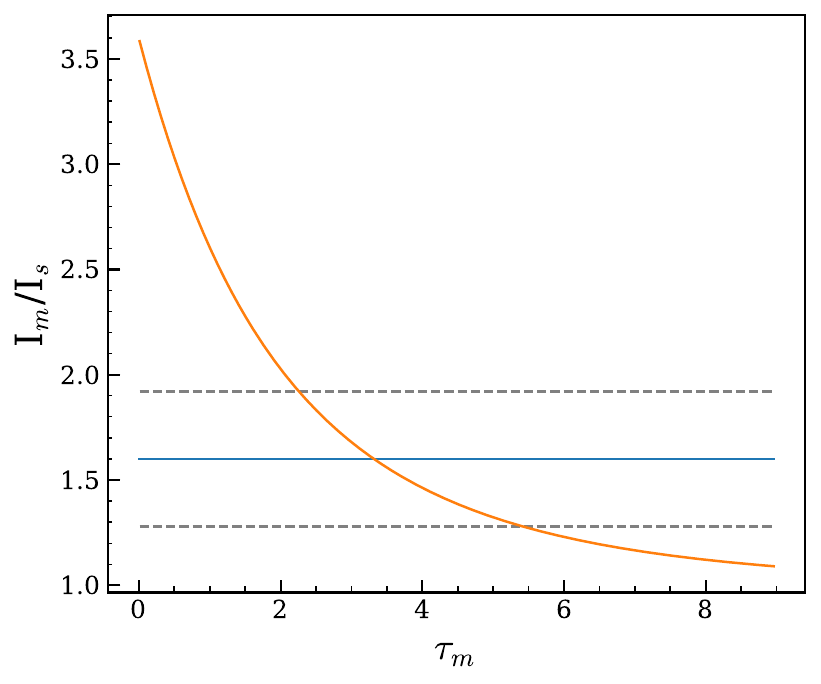}
    \includegraphics[width=0.3\linewidth]{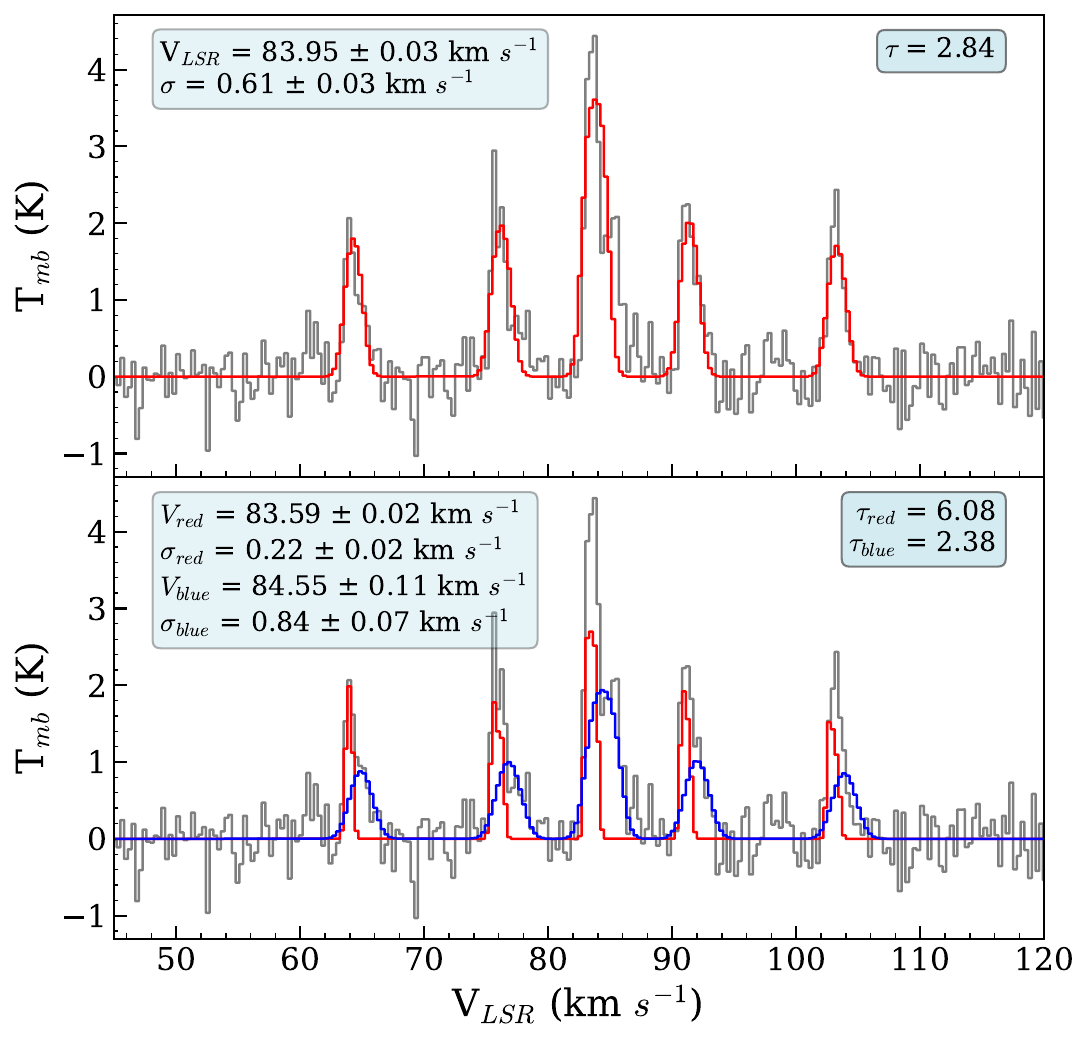}
    \caption{
The first panel shows the distribution of the NH$_3$(1,1) main line to inner left satellite line ratios in pixel space for AGAL022.376+00.447. 
The second panel presents the variation of this ratio as a function of optical depth for the NH$_3$(1,1) main line, where the yellow curve denotes the modeled trend, the blue horizontal line marks the reference ratio of 1.6, and the dashed lines indicate the corresponding 20\% uncertainty. 
The third panel displays the averaged NH$_3$(1,1) spectra of Core~1 in AGAL022.376+00.447, with the fitted components shown as blue and red profile.
}
\label{fig:core1}
\end{figure*}

In AGAL014.492–00.139, all single-component ammonia cores exhibit fully supersonic motions ($\mathcal{M} > 2$), indicating that nonthermal motions dominate the internal support. Outflows are detected in Cores~4 and~5 in the CO~$J=2\!-\!1$ transition, corresponding to the MM1 and MM2 cores identified by \citet{2019ApJ...886..130L}. The  Core~4 shows a narrow component with $\sigma_{v}=0.43$~km\,s$^{-1}$ and T$_{k}=21.6$~K, and a broad component with $\sigma_{v}=1.2$~km\,s$^{-1}$ and T$_{k}=16.6$~K. Similarly, Core~5 exhibits a narrow component ($\sigma_{v}=0.37$~km\,s$^{-1}$, T$_{k}=22.2$~K) and a broad one ($\sigma_{v}=0.92$~km\,s$^{-1}$, T$_{k}=14.0$~K). The broad component of Core~4 has the largest linewidth among all ammonia cores and corresponds to the longest outflow dynamical timescale of $2.51\times10^{4}$~yr, while Core~5 drives the maximum mass outflow of 0.148~M$_\odot$ with the shortest timescale of $0.4\times10^{4}$~yr. The broad component of Core 5 may indicate that it is more strongly perturbed by the outflow. The two components differ by a factor of $\sim$7 in total ammonia column density, with a peak value of N$_{tot}=8.2\times10^{15}$~cm$^{-2}$ in Core~5. The narrow components trace more quiescent gas ($\sigma_{non}=0.4$ and 0.33~km\,s$^{-1}$), although higher-density tracers such as N$_2$H$^+$ will be needed to confirm embedded cores \citep{2023MNRAS.526.4952P}.

In AGAL022.376+00.447, Core~2 shows a relatively low nonthermal velocity dispersion (\nonV\ = 0.36~km\,s$^{-1}$), while the other six cores yield Mach numbers of $\mathcal{M}=2.0$$\sim$3.4. Although most cores are supersonic, high-velocity CO~$J=2\!-\!1$ emission is not detected around them, except for Core~6, which exhibits a redshifted component offset by $\sim$0.1~pc from the peak of the 1.3~mm continuum. These results suggest that AGAL022.376+00.447 is at an earlier evolutionary stage than AGAL014.492–00.139, with less developed feedback activity.

AGAL024.314+00.086 shows a centrally concentrated morphology in both ammonia and 1.3~mm continuum maps, consisting of two clumps separated by $\Delta V_{\rm LSR}\approx3$~km\,s$^{-1}$. Outflows are detected in Cores~8 and~9. Core~9 is supersonic and has the highest ammonia column density (N$_{tot}\sim3.0\times10^{15}$~cm$^{-2}$) among all cores, consistent with the strong  outflow ($0.087$~M$_\odot$) associated with MM1 in \citet{2019ApJ...886..130L}. Core~8 contains both narrow and broad components: the narrow component ($\sigma_v = 0.25$~km\,s$^{-1}$) lies near the sound speed, whereas the broad component is supersonic, warmer (T$_{k}=19.3$~K), and likely perturbed and heated by the outflow.

The source AGAL031.024+00.262, located in the Scutum–Centaurus arm, contains two velocity components separated by $\sim$18~km\,s$^{-1}$. Core~18, which drives the outflow, exhibits two supersonic components, similar to Cores~16, 17, and~26, which are associated with 1.3~cm continuum emission. Feedback from embedded protostars appears to influence the surrounding gas over timescales longer than $\sim3.8\times10^{3}$~yr, comparable to the MM1 core in AGAL014.492–00.139 \citep{2019ApJ...886..130L}. The blue components at 78~km\,s$^{-1}$ trace nearly sonic gas in Cores~1, 24, and~25, while the red components at 96 \kms\ are subsonic in Cores~16 and~19, and approximately sonic in Core~21. The differences in the kinematic properties between the blue and red components suggest that they do not represent the same physical structure. Future observations using higher-density tracers such as N$_2$H$^+$  will be essential to clarify the evolutionary links between the components and to confirm the nature of the embedded protostellar activity \citep{2024MNRAS.530.1311F}.

\subsection{Optical depth effects on the extracted structure}
\label{sect:tau}

The NH$_3$(1,1) inversion transition consists of 18 hyperfine components arising from six underlying fine-structure levels, which are grouped into five observable spectral features comprising one main line and four satellite lines, including two inner and two outer satellites.
Under the assumption of local thermodynamic equilibrium (LTE) and optically thin conditions, the theoretical relative intensities of the inner satellite group is approximately 0.28, and those of the outer satellite group is approximately 0.22, with respect to the main line \citep[e.g.,][]{1983ARA&A..21..239H,2015PASP..127..266M}. 
This hyperfine structure allows an estimate of the optical depth $\tau$ under the LTE assumption, following equation~D1$\sim$D5 in \cite{2015PASP..127..266M}. The NH$_3$ column density is then corrected by the factor $\tau / (1 - e^{-\tau})$ \citep{2015PASP..127..266M}.

In AGAL022.376+00.447, the five NH$_3$(1,1) components observed with the VLA and GBT show predominantly single velocity components, enabling reliable measurement of the line intensity ratios. In contrast, the satellite lines in the other three clumps are strongly blended due to multiple velocity components. Therefore, the optical depths of the main component, $\tau_{\rm m}$, and the inner satellite component, $\tau_{\rm s}$, in AGAL022.376+00.447 are derived from the intensity ratio ($I_{\rm m}/I_{\rm s}$) between the main line and the inner satellite group, adopting their theoretical relative strengths.

The resulting histogram of $\tau_{\rm m}$ shows that most pixels fall within an $I_{\rm m}/I_{\rm s}$ ratio of 1.6--1.7.  Considering a 20\% uncertainty in this ratio, an observed value of $\sim$1.6 corresponds to $\tau_{\rm m} \approx 3.31$, with a minimum of 2.26 and a maximum of 5.4. The bias in $\tau_{\rm m}$ therefore roughly doubles and becomes further amplified in the column-density correction factor.
 The inner-satellite lines become optically thick at $\tau_{\rm m} \sim 3.6$, and all satellite lines become optically thick when $\tau_{\rm m} > 4.5$. Under such conditions, the main-line intensity becomes saturated, producing a flattened intensity distribution in the NH$_3$(1,1) main line. This effect may lead to an apparently uniform structure in regions where the emission is actually optically thick.

The presence of multiple velocity components affects the inferred optical depth, as illustrated in Fig~\ref{fig:core1}. Although the multi-component fit yields a smaller residual (RMS$_2$/RMS$_1$ = 0.88), the narrow component at lower velocities has an optical depth comparable to that obtained from a single component fit, and its satellite lines remain close to the optically thin regime. In contrast, the broad component is fully optically thick and likely traces denser gas regions.
Therefore, accurately resolving the structure requires higher angular resolution observations to distinguish between genuine substructures and self-absorption effects caused by optically thick NH$_3$(1,1) emission. However, for AGAL022.376+00.447, optical depths lower than $\tau_{\rm m} < 4.5$ have only a minor influence on the structural identification derived from the NH$_3$(1,1) satellite lines. We then corrected the intensity map of inner satellite group by the factor of  $\tau_s / (1 - e^{-\tau_s})$, as shown in Fig. \ref{fig:fig1}b.

\section{Summary}
\label{sec:col}

We presented the ammonia kinematics of four 70~$\mu$m dark molecular clumps based on VLA and GBT data. We used physical parameters derived from spectral line fitting of the NH$_3$(1,1) and NH$_3$(2,2) transitions to study the properties of ammonia cores on a scale of $\sim$0.1~pc.  The main conclusions are summarized as follows:

\begin{enumerate}
    \item We transformed the 1.3~cm continuum flux densities from the VLA observations into equivalent 1.3~mm intensities. The offsets between the 1.3~cm and 1.3~mm continuum indicate that the 1.3~cm emission does not originate from dust thermal emission. Moreover, the low value of $N_{\rm uv} \sim 10^{44}$ s$^{-1}$, compared with values reported in previous studies, suggests that the 1.3~cm continuum emission is not associated with H\,\textsc{ii} regions. These results indicate that the four clumps are primarily in an early evolutionary stage.

    \item Using the intensity of the NH$_3$(1,1) inner satellite group, we identified 92 ammonia cores across the four clumps with the \texttt{dendrogram} algorithm. Based on the optical-depth analysis of AGAL022.376+00.447, we determined that the NH$_3$(1,1) satellite structure is only weakly affected when $\tau_{\rm m} < 4.5$. Thirty-one small cores size of $<$ 2/3 beam were excluded from the analysis. Among the remaining 61 cores, 38 exhibit single velocity components and 23 show two components in the NH$_3$(1,1) main line. Twelve ammonia cores are associated with 1.3~mm continuum emission, and five are associated with the outflows of CO J=2–1 line.

    \item For the 38 single-component ammonia cores, the velocity dispersion $\sigma_v$  ranges from 0.23~km\,s$^{-1}$ on Core~1 of AGAL031.024+00.262  to 1.20~km\,s$^{-1}$ on Core~13 of AGAL024.314+00.086. The kinetic temperature T$_{k}$ spans 11.6 $\sim$16.5~K. To investigate the internal kinematics of the ammonia cores, we averaged the spectra of 61 ammonia cores within concentric annuli and derived the nonthermal components from the NH$_3$(1,1) main line in each annulus. We find a clear trend in which nonthermal motions of supersonic or sonic increase with radius in five ammonia cores of AGAL022.376+00.447 and AGAL024.314+00.086. In AGAL031.024+00.262, four cores show a similar outward increase, while seven cores display the opposite trend. The former ammonia cores likely experience turbulence dissipation and are approaching gravitational collapse. In contrast, seven cores with decreasing nonthermal components toward outer regions are likely influenced by feedback from ongoing star formation activity.
\end{enumerate}

\begin{acknowledgements}
    We thank the referee for the comments and suggestions that improved this work.
    We sincerely thank  Prof. Zhiyu Zhang, Lingrui Lin, and Yichen Sun for their help with data reduction. We also acknowledge helpful with Chang Ruan, Rui Luo, Xing Lu, Fengyao Zhu, Siqi Zheng, and Fei Li. S.L. acknowledges support from the National SKA Program of China with No. 2025SKA0140100, “Double First-Class” Funding with No. 14912217, and National Natural Science Foundation of China (NSFC) grant with No. 13004007. This work is supported by National Key R$\&$D Program of China under grant 2023YFA1608204, the National Natural Science Foundation of China grant 12550003 and the Guangxi Talent Programme (Highland of Innovation Talents). 

\end{acknowledgements}

\bibliographystyle{aa}
\bibliography{ref}

\begin{appendix}

\onecolumn
\section{Physical parameters of ammonia cores derived from \amm\ line fitting}
\setlength{\tabcolsep}{3pt}
\begin{longtable}{ccccccccccccc}
\caption{The averaged parameters for total ammonia cores.}\\
\label{tab:tab3} \\
\hline\hline
ID & R.A. & Decl. & Maj & Min& PA & V$_{LSR}$ & T$_{k}$ & N$_{tot}$ & $\sigma_{v}$ & $\sigma_{th}$ & $\sigma_{non}$ & $\mathcal{M}$ \\
Num. & (hh:mm:ss) & (dd:mm:ss) & ($\arcsec$) & ($\arcsec{}$) & ($^{\circ}$) & (km s$^{-1}$) & (K) & (10$^{14}$ cm$^{-2}$) & (km s$^{-1}$) & (km s$^{-1}$) & (km s$^{-1}$) & \\
\hline
\endfirsthead
\caption{continued.}\\
\hline\hline
ID & R.A. & Decl. & Maj & Min& PA & V$_{LSR}$ & T$_{k}$ & N$_{tot}$ & $\sigma_{v}$ & $\sigma_{th}$ & $\sigma_{non}$ & $\mathcal{M}$ \\
Num. & (hh:mm:ss) & (dd:mm:ss) & ($\arcsec$) & ($\arcsec{}$) & ($^{\circ}$) & (km s$^{-1}$) & (K) & (10$^{14}$ cm$^{-2}$) & (km s$^{-1}$) & (km s$^{-1}$) & (km s$^{-1}$) & \\
\hline
\endhead
\hline
\endfoot

\endlastfoot

& & & & & \multicolumn{3}{c}{\textbf{AGAL014.492–00.139}} & & & & & \\
\hline
1 & 18:17:24.22 & -16:25:32.44 & 10.0 & 4.5 & -152 & 39.7(0.03) & 15.25(0.77) & 14.79(1.19) & 0.71(0.03) & 0.23 & 0.7 & 3.0 \\
2 & 18:17:24.23 & -16:25:24.61 & 3.1 & 2.4 & 64 & 39.71(0.05) & 15.84(1.21) & 11.07(1.48) & 0.73(0.04) & 0.24 & 0.71 & 3.0 \\
3 & 18:17:22.77 & -16:25:17.35 & 6.7 & 4.8 & 93 & 38.37(0.02) & 16.03(0.86) & 17.54(1.37) & 0.51(0.02) & 0.24 & 0.48 & 2.03 \\
4 & 18:17:21.50 & -16:25:02.22 & 3.9 & 2.3 & 164 & 40.01(0.07) & 16.64(0.71) & 46.24(2.87) & 1.2(0.04) & 0.24 & 1.19 & 4.92 \\
& & & & & & 41.5(0.03) & 21.59(2.11) & 14.72(2.03) & 0.43(0.03) & 0.28 & 0.4 & 1.45 \\
5 & 18:17:22.33 & -16:24:58.48 & 6.0 & 4.1 & -168 & 39.42(0.04) & 13.96(0.43) & 81.85(4.71) & 0.92(0.03) & 0.22 & 0.9 & 4.06 \\
& & & & & & 41.11(0.03) & 22.19(2.14) & 11.53(1.75) & 0.37(0.03) & 0.28 & 0.33 & 1.18 \\
6 & 18:17:21.34 & -16:24:49.61 & 4.6 & 4.3 & 58 & 40.24(0.02) & 15.12(0.39) & 43.25(1.59) & 0.9(0.02) & 0.23 & 0.89 & 3.85 \\
7 & 18:17:25.25 & -16:25:17.67 & 3.4 & 2.4 & 139 & 39.73(0.07) & 12.59(0.9) & 37.15(4.79) & 1.1(0.06) & 0.21 & 1.09 & 5.18 \\
8 & 18:17:25.14 & -16:25:10.71 & 5.1 & 2.1 & 97 & 39.3(0.05) & 13.3(0.82) & 21.28(1.96) & 0.93(0.04) & 0.22 & 0.91 & 4.21 \\
9 & 18:17:24.51 & -16:25:06.22 & 6.1 & 3.4 & 69 & 38.66(0.04) & 17.53(3.38) & 6.08(1.89) & 0.25(0.04) & 0.25 & 0.2 & 0.8 \\
& & & & & & 39.41(0.11) & 13.8(1.32) & 38.99(6.64) & 1.43(0.08) & 0.22 & 1.42 & 6.41 \\
10 & 18:17:20.00 & -16:24:54.49 & 4.0 & 3.1 & 50 & 40.11(0.05) & 15.29(1.02) & 21.63(2.34) & 0.85(0.04) & 0.23 & 0.84 & 3.61 \\
11 & 18:17:24.01 & -16:24:49.58 & 6.5 & 4.0 & 69 & 38.16(0.06) & 13.72(2.46) & 7.94(2.38) & 0.37(0.06) & 0.22 & 0.33 & 1.52 \\
& & & & & & 40.49(0.17) & 15.33(2.38) & 54.45(18.81) & 1.01(0.15) & 0.23 & 1.0 & 4.27 \\
\hline
\multicolumn{6}{r}{Median} & 39.71 & 15.29 & 21.28 & 0.85 & 0.23 & 0.84 & 3.61 \\
\multicolumn{6}{r}{Mean} & 39.73 & 15.88 & 28.57 & 0.78 & 0.24 & 0.76 & 3.30 \\
\hline
& & & & &  \multicolumn{3}{c}{\textbf{AGAL022.376+00.447}} &  & & & & \\
\hline
1 & 18:30:36.46 & -09:13:18.14 & 10.4 & 5.1 & -155 & 84.04(0.02) & 12.26(0.53) & 25.7(1.66) & 0.57(0.02) & 0.21 & 0.56 & 2.66 \\
2 & 18:30:36.95 & -09:13:07.16 & 9.0 & 3.9 & 94 & 84.82(0.02) & 12.25(0.56) & 17.99(1.12) & 0.39(0.02) & 0.21 & 0.36 & 1.75 \\
3 & 18:30:37.25 & -09:13:00.26 & 7.8 & 2.1 & 53 & 85.0(0.02) & 12.32(0.72) & 17.34(1.4) & 0.45(0.02) & 0.21 & 0.42 & 2.01 \\
4 & 18:30:37.51 & -09:12:47.84 & 3.7 & 2.2 & -176 & 84.71(0.03) & 15.71(0.82) & 27.48(2.09) & 0.64(0.03) & 0.24 & 0.63 & 2.65 \\
5 & 18:30:37.65 & -09:12:41.18 & 7.2 & 2.9 & 53 & 84.31(0.03) & 13.74(0.53) & 32.06(1.77) & 0.76(0.02) & 0.22 & 0.74 & 3.36 \\
6 & 18:30:38.23 & -09:12:44.91 & 3.5 & 1.9 & -157 & 84.52(0.03) & 14.05(0.78) & 23.12(1.76) & 0.68(0.03) & 0.22 & 0.66 & 2.95 \\
7 & 18:30:36.60 & -09:12:42.53 & 3.9 & 2.3 & 73 & 84.22(0.05) & 14.89(1.21) & 16.07(2.04) & 0.79(0.05) & 0.23 & 0.77 & 3.37 \\
\hline
\multicolumn{6}{r}{Median} & 84.52 & 13.74 & 23.12 & 0.64 & 0.22 & 0.63 & 2.66 \\
\multicolumn{6}{r}{Mean} & 84.52 & 13.60 & 22.82 & 0.61 & 0.22 & 0.59 & 2.68 \\
\hline
& & & & &  \multicolumn{3}{c}{\textbf{AGAL024.314+00.086}} & & & & & \\
\hline
1 & 18:35:16.83 & -07:37:38.09 & 8.9 & 6.3 & -155 & 113.97(0.2) & ... & ...& ... & ... & ... & ... \\
2 & 18:35:15.90 & -07:37:28.59 & 9.5 & 5.0 & 81 & 113.92(0.03) & 16.57(0.91) & 8.61(0.85) & 0.57(0.03) & 0.24 & 0.55 & 2.27 \\
& & & & & & 117.88(0.07) & 14.02(2.39) & 10.45(3.39) & 0.43(0.06) & 0.22 & 0.41 & 1.82 \\
3 & 18:35:16.38 & -07:37:01.21 & 10.5 & 4.0 & 81 & 115.15(0.17) & 12.89(1.91) & 7.33(3.26) & 1.84(0.14) & 0.21 & 1.84 & 8.61 \\
& & & & & & 116.33(0.03) & 13.3(1.91) & 5.02(1.23) & 0.18(0.02) & 0.22 & 0.1 & 0.46 \\
4 & 18:35:18.04 & -07:37:03.09 & 6.0 & 2.3 & 78 & 114.95(0.04) & 15.12(1.02) & 10.5(1.18) & 0.57(0.03) & 0.23 & 0.55 & 2.39 \\
& & & & & & 117.69(0.11) & 26.61(6.27) & 4.16(2.95) & 0.58(0.12) & 0.31 & 0.56 & 1.82 \\
5 & 18:35:20.48 & -07:37:12.12 & 6.3 & 4.6 & 163 & 113.67(0.03) & 13.58(0.62) & 12.11(0.84) & 0.74(0.02) & 0.22 & 0.68 & 3.12 \\
6 & 18:35:20.89 & -07:37:05.43 & 5.7 & 2.8 & 79 & 114.3(0.01) & 15.09(0.54) & 15.96(0.88) & 0.44(0.01) & 0.23 & 0.41 & 1.77 \\
7 & 18:35:18.62 & -07:37:51.53 & 4.3 & 3.3 & 172 & 116.74(0.01) & 13.91(0.46) & 13.3(0.61) & 0.36(0.01) & 0.22 & 0.33 & 1.47 \\
8 & 18:35:18.64 & -07:37:36.90 & 7.0 & 2.5 & 88 & 115.87(0.1) & 19.25(1.56) & 16.71(2.73) & 1.44(0.08) & 0.26 & 1.43 & 5.47 \\
& & & & & & 116.76(0.01) & 11.67(0.55) & 15.78(1.13) & 0.25(0.01) & 0.2 & 0.21 & 1.01 \\
9 & 18:35:19.18 & -07:37:22.64 & 13.1 & 8.3 & -162 & 114.18(0.02) & 14.57(0.33) & 30.34(0.98) & 0.88(0.02) & 0.23 & 0.87 & 3.83 \\
10 & 18:35:18.33 & -07:37:22.23 & 8.9 & 2.5 & 75 & 114.62(0.02) & 15.36(0.63) & 14.39(0.89) & 0.6(0.02) & 0.23 & 0.58 & 2.47 \\
& & & & & & 117.34(0.02) & 14.14(0.76) & 7.67(0.62) & 0.32(0.02) & 0.22 & 0.29 & 1.28 \\
11 & 18:35:19.42 & -07:37:07.85 & 6.8 & 3.9 & 102 & 114.88(0.03) & 15.16(0.66) & 18.32(1.27) & 0.9(0.03) & 0.23 & 0.88 & 3.8 \\
12 & 18:35:20.23 & -07:36:55.11 & 6.9 & 3.0 & 59 & 114.49(0.02) & 15.23(0.69) & 11.12(0.79) & 0.57(0.02) & 0.23 & 0.55 & 2.38 \\
13 & 18:35:19.66 & -07:36:52.25 & 4.4 & 2.5 & 70 & 115.14(0.07) & 16.54(1.19) & 17.66(2.32) & 1.21(0.06) & 0.24 & 1.2 & 4.94 \\
14 & 18:35:21.46 & -07:36:56.71 & 6.3 & 3.3 & 100 & 113.41(0.02) & 15.44(0.71) & 11.04(0.81) & 0.55(0.02) & 0.23 & 0.53 & 2.25 \\
15 & 18:35:22.15 & -07:37:10.37 & 9.8 & 6.3 & -142 & 113.3(0.01) & 15.11(0.51) & 14.55(0.77) & 0.4(0.01) & 0.23 & 0.37 & 1.58 \\
\hline
\multicolumn{6}{r}{Median} & 114.91 & 15.11 & 12.71 & 0.57 & 0.23 & 0.55 & 2.33 \\
\multicolumn{6}{r}{Mean} & 115.23 & 15.37 & 13.30 & 0.74 & 0.23 & 0.72 & 3.08 \\
\hline
& & & & &  \multicolumn{3}{c}{\textbf{AGAL031.024+00.262}} &  & & & & \\
\hline
1 & 18:47:02.09 & -01:35:32.02 & 4.3 & 2.5 & 125 & 78.62(0.02) & ... & ... & 0.27(0.02) & 0.21 & 0.22 & 1.05 \\
2 & 18:47:02.80 & -01:35:26.06 & 5.7 & 2.5 & -178 & 78.63(0.02) & ... & ... & 0.29(0.02) & 0.21 & 0.25 & 1.16 \\
3 & 18:47:04.10 & -01:35:20.83 & 4.9 & 2.5 & 52 & 78.7(0.02) & ... & ... & 0.3(0.02) & 0.21 & 0.26 & 1.21 \\
4 & 18:47:04.49 & -01:35:18.73 & 3.0 & 2.1 & -171 & 78.72(0.02) & ... & ... & 0.34(0.02) & 0.21 & 0.31 & 1.42 \\
5 & 18:47:03.32 & -01:35:12.12 & 4.5 & 2.1 & 127 & 78.25(0.02) & ... & ... & 0.33(0.02) & 0.21 & 0.29 & 1.37 \\
6 & 18:47:04.10 & -01:35:09.10 & 4.4 & 2.8 & -159 & 78.57(0.05) & ... & ... & 0.63(0.05) & 0.21 & 0.61 & 2.85 \\
7 & 18:47:02.80 & -01:35:10.91 & 3.5 & 2.5 & 104 & 78.41(0.03) & 12.71(1.02) & 5.89(0.68) & 0.44(0.03) & 0.21 & 0.41 & 1.95 \\
8 & 18:47:02.65 & -01:35:05.85 & 6.4 & 2.8 & 152 & 78.13(0.03) & ... & ... & 0.53(0.04) & 0.21 & 0.51 & 2.37 \\
9 & 18:47:01.21 & -01:35:05.65 & 4.5 & 2.7 & 117 & 78.69(0.05) & ... & ... & 0.54(0.05) & 0.21 & 0.52 & 2.42 \\
10 & 18:47:03.32 & -01:34:51.02 & 3.2 & 2.5 & 164 & 78.5(0.03) & 13.74(1.31) & 7.94(1.28) & 0.34(0.03) & 0.22 & 0.3 & 1.38 \\
11 & 18:47:03.13 & -01:34:40.63 & 5.2 & 2.7 & 116 & 78.4(0.04) & ... & ... & 0.52(0.03) & 0.21 & 0.5 & 2.32 \\
12 & 18:47:02.50 & -01:34:38.95 & 3.9 & 2.4 & -162 & 78.44(0.05) & ... & ... & 0.47(0.05) & 0.21 & 0.45 & 2.08 \\
13 & 18:47:02.44 & -01:34:32.63 & 3.6 & 2.5 & -155 & 78.15(0.03) & 11.62(1.17) & 5.37(0.74) & 0.46(0.03) & 0.2 & 0.44 & 2.15 \\
15 & 18:46:59.49 & -01:34:17.48 & 4.2 & 1.5 & 168 & 95.48(0.01) & ... & ... & 0.31(0.01) & 0.21 & 0.27 & 1.26 \\
16 & 18:47:01.63 & -01:35:01.21 & 4.3 & 2.1 & -141 & 78.67(0.02) & ... & ... & 0.41(0.02) & 0.21 & 0.39 & 1.8 \\
& & & & & & 96.84(0.04) & ... & ... & 0.18(0.03) & 0.21 & 0.09 & 0.43 \\
17 & 18:47:01.01 & -01:34:42.79 & 5.5 & 4.2 & 96 & 78.59(0.03) & 12.82(1.02) & 7.13(0.9) & 0.54(0.03) & 0.21 & 0.52 & 2.44 \\
& & & & & & 96.18(0.02) & 14.56(0.68) & 10.64(0.78) & 0.55(0.02) & 0.23 & 0.53 & 2.34 \\
18 & 18:47:01.72 & -01:34:44.05 & 8.9 & 6.3 & 81 & 78.15(0.01) & 11.3(0.5) & 7.08(0.37) & 0.43(0.01) & 0.2 & 0.4 & 2.0 \\
& & & & & & 96.38(0.01) & 13.04(0.32) & 12.25(0.39) & 0.47(0.01) & 0.21 & 0.45 & 2.09 \\
19 & 18:47:01.72 & -01:34:30.44 & 3.1 & 1.6 & 118 & 78.15(0.05) & ... & ... & 0.44(0.04) & 0.21 & 0.41 & 1.91 \\
& & & & & & 96.17(0.01) & 10.83(0.66) & 7.85(0.58) & 0.22(0.01) & 0.2 & 0.16 & 0.82 \\
20 & 18:47:01.38 & -01:34:25.18 & 3.8 & 2.0 & 139 & 78.24(0.06) & ... & ... & 0.39(0.07) & 0.21 & 0.36 & 1.68 \\
& & & & & & 96.19(0.02) & 12.13(0.43) & 16.03(0.78) & 0.48(0.01) & 0.21 & 0.45 & 2.19 \\
21 & 18:47:02.33 & -01:34:18.68 & 4.4 & 2.4 & 65 & 78.13(0.07) & ... & ... & 0.6(0.06) & 0.21 & 0.58 & 2.72 \\
& & & & & & 95.58(0.02) & 10.58(0.89) & 5.93(0.56) & 0.26(0.02) & 0.19 & 0.22 & 1.12 \\
22 & 18:47:00.24 & -01:34:18.95 & 7.2 & 4.2 & -151 & 77.59(0.04) & 12.92(1.36) & 3.07(0.85) & 0.58(0.04) & 0.21 & 0.56 & 2.62 \\
& & & & & & 95.64(0.03) & 15.6(0.78) & 7.03(0.79) & 0.78(0.03) & 0.24 & 0.76 & 3.24 \\
23 & 18:47:01.08 & -01:34:18.25 & 2.9 & 1.2 & 57 & 78.04(0.07) & ... & ... & 0.72(0.08) & 0.21 & 0.71 & 3.31 \\
& & & & & & 96.37(0.05) & ... & ... & 0.45(0.04) & 0.21 & 0.42 & 1.98 \\
24 & 18:47:00.52 & -01:34:12.34 & 3.3 & 1.4 & -155 & 77.14(0.02) & 10.53(0.88) & 7.74(0.96) & 0.18(0.01) & 0.19 & 0.11 & 0.58 \\
& & & & & & 96.21(0.05) & 14.07(1.27) & 6.17(1.18) & 0.66(0.05) & 0.22 & 0.64 & 2.89 \\
25 & 18:47:02.01 & -01:34:09.42 & 4.4 & 3.0 & 66 & 78.46(0.02) & ... & ... & 0.24(0.03) & 0.21 & 0.19 & 0.87 \\
& & & & & & 95.9(0.02) & 15.11(0.84) & 5.87(0.61) & 0.45(0.02) & 0.23 & 0.42 & 1.82 \\
26 & 18:47:01.65 & -01:34:08.95 & 3.1 & 1.4 & 75 & 78.19(0.05) & ... & ... & 0.35(0.05) & 0.21 & 0.32 & 1.49 \\
& & & & & & 95.69(0.05) & ... & ... & 0.65(0.05) & 0.21 & 0.63 & 2.93 \\
27 & 18:47:02.43 & -01:34:05.60 & 5.1 & 2.4 & 60 & 78.7(0.04) & ... & ... & 0.29(0.04) & 0.21 & 0.25 & 1.17 \\
& & & & & & 95.85(0.02) & 12.36(0.75) & 7.5(0.64) & 0.42(0.02) & 0.21 & 0.4 & 1.89 \\
28 & 18:46:59.76 & -01:34:10.00 & 4.8 & 1.7 & 107 & 95.87(0.09) & ... & ... & 0.9(0.08) & 0.21 & 0.89 & 4.14 \\
\hline

\multicolumn{6}{r}{Median$_{blue}$} & 78.41 & 12.71 & 7.08 & 0.43 & 0.21 & 0.40 & 1.91 \\
\multicolumn{6}{r}{Mean$_{blue}$} & 76.51 & 12.23 & 6.32 & 0.43 & 0.21 & 0.39 & 1.85 \\
\multicolumn{6}{r}{Median$_{red}$} & 96.03 & 13.04 & 7.50 & 0.46 & 0.21 & 0.43 & 2.04 \\
\multicolumn{6}{r}{Mean$_{red}$} & 96.02 & 13.14 & 8.81 & 0.48 & 0.21 & 0.45 & 2.08 \\
\hline
\\
\multicolumn{13}{p{\textwidth}}{\footnotesize
  \textbf{Note.}
  The core IDs of the ammonia cores in each source are listed in Column~1.
  Columns~2--6 give the central coordinates and size parameters of the ammonia
  cores. Columns~7--13 present the physical parameters derived from the line
  fitting and subsequent calculations, as described in Sect.~\ref{sec:3.2.2}.
  The median and mean values of the parameters are listed in the last two rows
  for each source.
  } \\

\end{longtable}

\newpage
\section{Radial variations in the properties of ammonia cores}
\setlength{\tabcolsep}{6pt}
\begin{longtable}{ccccccc}
\caption{Radial variations of kinetic temperature and nonthermal velocity dispersion in ammonia cores, with 1.3~mm continuum and
  CO~2--1 outflow associations.}\\
\label{tab:tab4} \\
\hline\hline
Core ID & R$_{an}$ & T$_k$ & $\sigma_{non}$ & $\Delta\sigma_{non}$ & 1.3 mm & Outflow \\
Num. & pc & (K) & (km s$^{-1}$) & (km s$^{-1}$) &  &  \\
\hline

\endfirsthead
\caption{continued.}\\
\hline\hline
Core ID & R$_{an}$ & T$_k$ & $\sigma_{non}$ & $\Delta\sigma_{non}$ & 1.3 mm & Outflow \\
Num. & pc & (K) & (km s$^{-1}$) & (km s$^{-1}$) &  &  \\
\hline
\endhead
\hline
\endfoot

\endlastfoot
\hline
\multicolumn{7}{c}{\textbf{AGAL014.492–00.139}} \\  
\hline

1 & 0.0072$\sim$0.0495 & 15.36(0.81)$\sim$15.73(0.56) & 0.66(0.03)$\sim$0.67(0.02) & 0.01 &   &   \\
\hline
\multicolumn{7}{c}{\textbf{AGAL022.376+00.447}} \\ 
\hline
1 & 0.0102$\sim$0.071 & 11.54(0.62)$\sim$13.25(0.39) & 0.54(0.02)$\sim$0.59(0.01) & 0.05 &   &   \\
2 & 0.0102$\sim$0.071 & 11.84(0.64)$\sim$13.16(0.45) & 0.34(0.02)$\sim$0.41(0.01) & 0.07 & \checkmark &   \\
\hline
\multicolumn{7}{c}{\textbf{AGAL024.314+00.086}} \\ 
\hline
9 & 0.0157$\sim$0.1091 & 14.82(0.67)$\sim$15.47(0.8) & 0.64(0.01)$\sim$0.74(0.01) & 0.10 & \checkmark & \checkmark \\
11 & 0.0157$\sim$0.078 & 15.80(0.63)$\sim$17.5(0.66) & 0.71(0.02)$\sim$0.73(0.02) & 0.02 & \checkmark & \checkmark \\
12 & 0.0157$\sim$0.078 & 15.20(0.60)$\sim$14.30(0.51) & 0.53(0.02)$\sim$0.69(0.02) & 0.16 &   &   \\
15 & 0.0157$\sim$0.1091 & 24.75(5.69)$\sim$16.07(0.59) & 0.19(0.06)$\sim$0.43(0.01) & 0.24 &   &   \\
\hline
\multicolumn{7}{c}{\textbf{AGAL031.024+00.262}} \\  
\hline
1 & 0.0086$\sim$0.0423 & 11.91(1.21)$\sim$11.37(0.75) & 0.23(0.03)$\sim$0.39(0.02) & 0.16 &   &   \\
\\
6 & 0.0086$\sim$0.0423 & 12.42(1.51)$\sim$10.96(0.90) & 0.62(0.05)$\sim$0.48(0.02) & 0.14 &   &   \\
8 & 0.0086$\sim$0.0423 & 11.38(1.37)$\sim$11.95(0.76) & 0.54(0.04)$\sim$0.39(0.02) & 0.15 &   &   \\
9 & 0.0086$\sim$0.0423 & 10.76(1.63)$\sim$12.49(0.7) & 0.52(0.04)$\sim$0.32(0.02) & 0.20 &   &   \\
\\
17 & 0.0086$\sim$0.0423 & 12.68(0.96)$\sim$14.89(0.68) & 0.50(0.03)$\sim$0.52(0.02) & 0.02 & \checkmark & \checkmark \\
 & 0.0086$\sim$0.0423 & 14.84(0.73)$\sim$14.40(0.42) & 0.56(0.02)$\sim$0.41(0.01) & 0.15 &   &   \\
18 & 0.0086$\sim$0.0592 & 11.49(0.64)$\sim$12.08(0.43) & 0.39(0.02)$\sim$0.42(0.01) & 0.03 & \checkmark &   \\
 & 0.0086$\sim$0.0592 & 12.72(0.35)$\sim$14.74(0.31) & 0.48(0.01)$\sim$0.38(0.01) & 0.10 &   &   \\
21 & 0.0086$\sim$0.0423 & 12.92(1.47)$\sim$12.44(0.75) & 0.58(0.06)$\sim$0.35(0.02) & 0.23 &   &   \\
 & 0.0086$\sim$0.0423 & 10.38(0.73)$\sim$12.41(0.69) & 0.22(0.02)$\sim$0.36(0.02) & 0.14 &   &   \\
22 & 0.0086$\sim$0.0592 & 12.44(1.83)$\sim$12.87(0.99) & 0.57(0.04)$\sim$0.52(0.03) & 0.05 &   &   \\
 & 0.0086$\sim$0.0592 & 15.43(0.73)$\sim$14.51(0.37) & 0.74(0.03)$\sim$0.62(0.01) & 0.12 &   &   \\
\hline

\\
 \multicolumn{7}{p{\textwidth}}{\footnotesize
  \textbf{Note.}
  Column~1 lists the ammonia core IDs.
  Columns~2--4 give the radius, kinetic temperature, and nonthermal velocity
  dispersion from the central annulus and the outer annulus, respectively.
  Column~5 shows the variation in nonthermal velocity dispersion between the
  inner and outer annuli. Columns~6 and~7 indicate whether each core is associated
  with 1.3~mm continuum emission and outflow of CO~$J=2$--1, respectively.
  }
\end{longtable}
\onecolumn

\begin{figure}[ht!]
    \centering
    \includegraphics[width=0.5\linewidth]{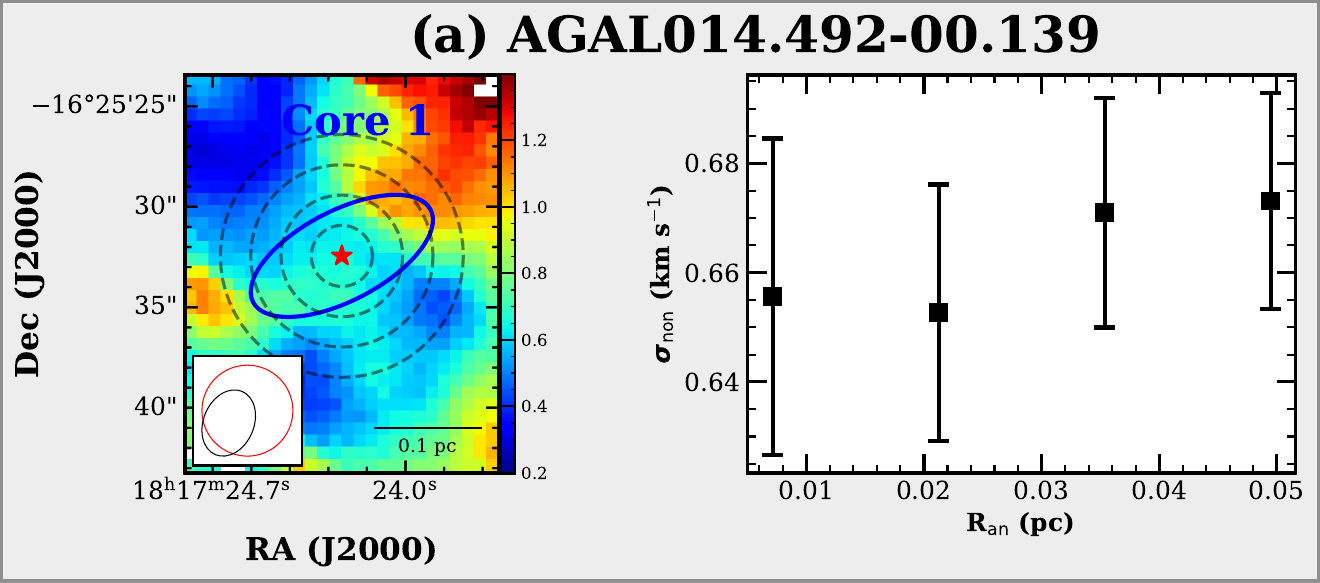}\\[1ex]
    \vspace{1em}
    \includegraphics[width=1\linewidth]{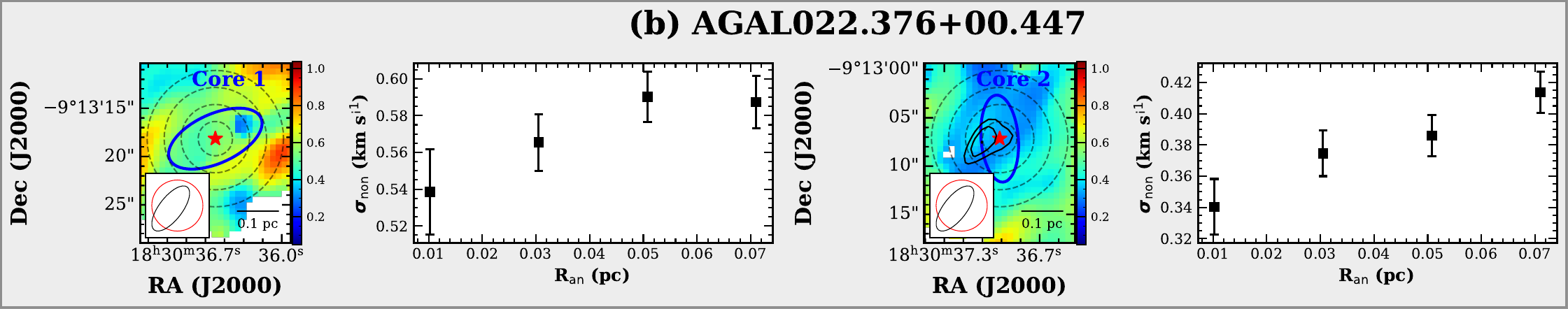}\\[1ex]
    \vspace{1em}
        \includegraphics[width=1\linewidth]{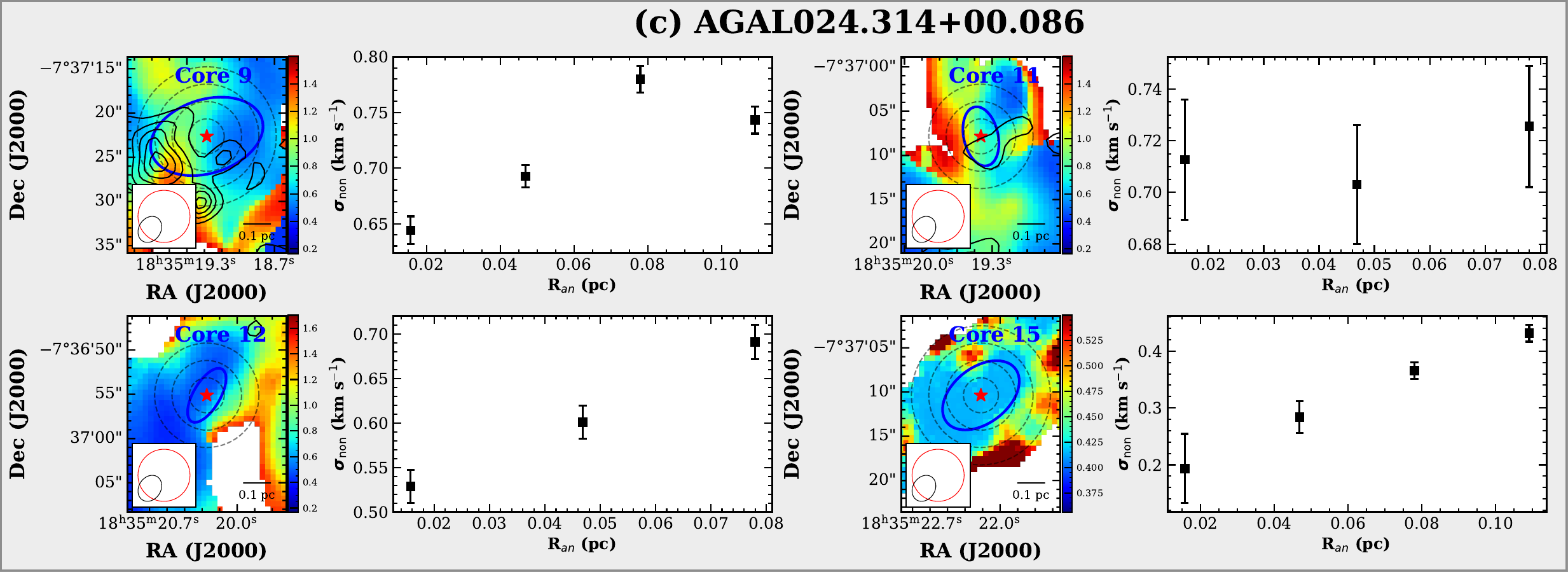}\\[1ex]
    \caption{The colored background in the left panel denotes the \nonV\ map of the NH$_{3}$(1,1) main line. The dashed circular ring outlines the averaged regions used for the NH$_{3}$(1,1) and NH$_{3}$(2,2) spectra. The black contours show the 1.3 mm continuum with the same contour levels as those in Fig. \ref{fig:fig1}.  The black circle indicates the beam size of the 1.3 mm continuum observation, while the red circle denotes the beam size of the NH$_{3}$(1,1) line. The right panel displays the radial variation of \nonV\ from the center to the outer regions, marked in black. }
    \label{fig:figB}
\end{figure}

\end{appendix}
\end{document}